%% file: second_version_arXiv.tex
\documentclass[prd,superscriptaddress, reprint, floatfix]{revtex4-1}

\usepackage[english]{babel}
\usepackage{graphicx}
\usepackage{xcolor}
\usepackage{amssymb,amsmath, wasysym}
\usepackage{tikz}
\usepackage{times}
\usetikzlibrary{backgrounds}
\usetikzlibrary{arrows}
\usetikzlibrary{shapes}

\usepackage{siunitx}

\include{JournalAbbr}    

\begin{document}
\hfill{IPPP/19/58}
\title[Galaxy formation in modified gravity]
{Realistic simulations of galaxy formation in $f(R)$ modified gravity}

\author{Christian \surname{Arnold}}
\email{christian.arnold@durham.ac.uk}
\affiliation{Institute for Computational Cosmology, Department of Physics, Durham University, South Road, Durham DH13LE, UK}

\author{Matteo \surname{Leo}}
\affiliation{Institute for Computational Cosmology, Department of Physics, Durham University, South Road, Durham DH13LE, UK}
\affiliation{Institute for Particle Physics Phenomenology, Department of Physics, Durham University, South Road, Durham DH13LE, UK}

\author{Baojiu \surname{Li}}
\affiliation{Institute for Computational Cosmology, Department of Physics, Durham University, South Road, Durham DH13LE, UK}

\date{July 8, 2019 - preprint, published in Nature Astronomy}


\begin{abstract}
We have carried out a set of cosmological hydrodynamical simulations that follow galaxy formation in $f(R)$ modified gravity models. Our simulations employ the Illustris-TNG full physics model and a new modified gravity solver in the \textsc{arepo} code. For the first time we are able to investigate the degeneracy in the matter power spectrum between the effects of $f(R)$-gravity and feedback from active galactic nuclei (AGN), and the imprint of modified gravity on the properties of galaxies and on the distribution of dark matter, gas and stars in the universe. $f(R)$-gravity has an observable effect on the neutral hydrogen power spectrum at high redshift at a level of $20\%$. For both the F6 and F5 models, this is significantly larger than the predicted errors for the SKA1-MID survey, making this probe a powerful test of gravity on large scales. A similar effect is present in the power spectrum of the stars at high redshift. We also show that rotationally supported disc galaxies can form in $f(R)$-gravity, even in the partially screened regime. Our simulations indicate that there might be more disc galaxies in F6 compared to GR, and fewer in F5. Finally, we show that the back reaction between AGN feedback and modified gravity in the matter power spectrum is not important in the F6 model but has a sizeable effect in F5. 
\end{abstract}

\maketitle

\renewcommand{\d}{{\rm d}}

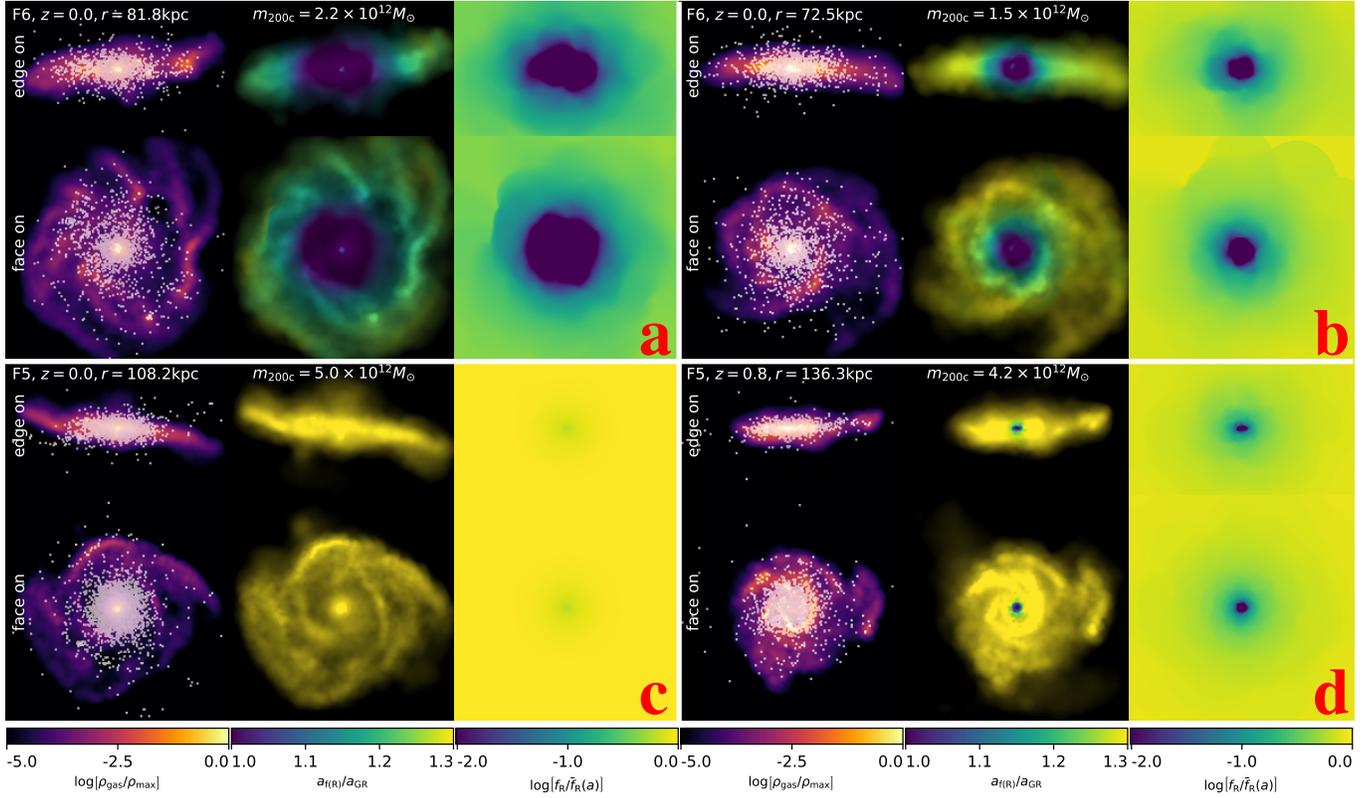
\begin{figure*}
\centering
\begin{tikzpicture}
\node[anchor = south west, inner sep = 0] at (0,0) {\includegraphics[width = \textwidth]{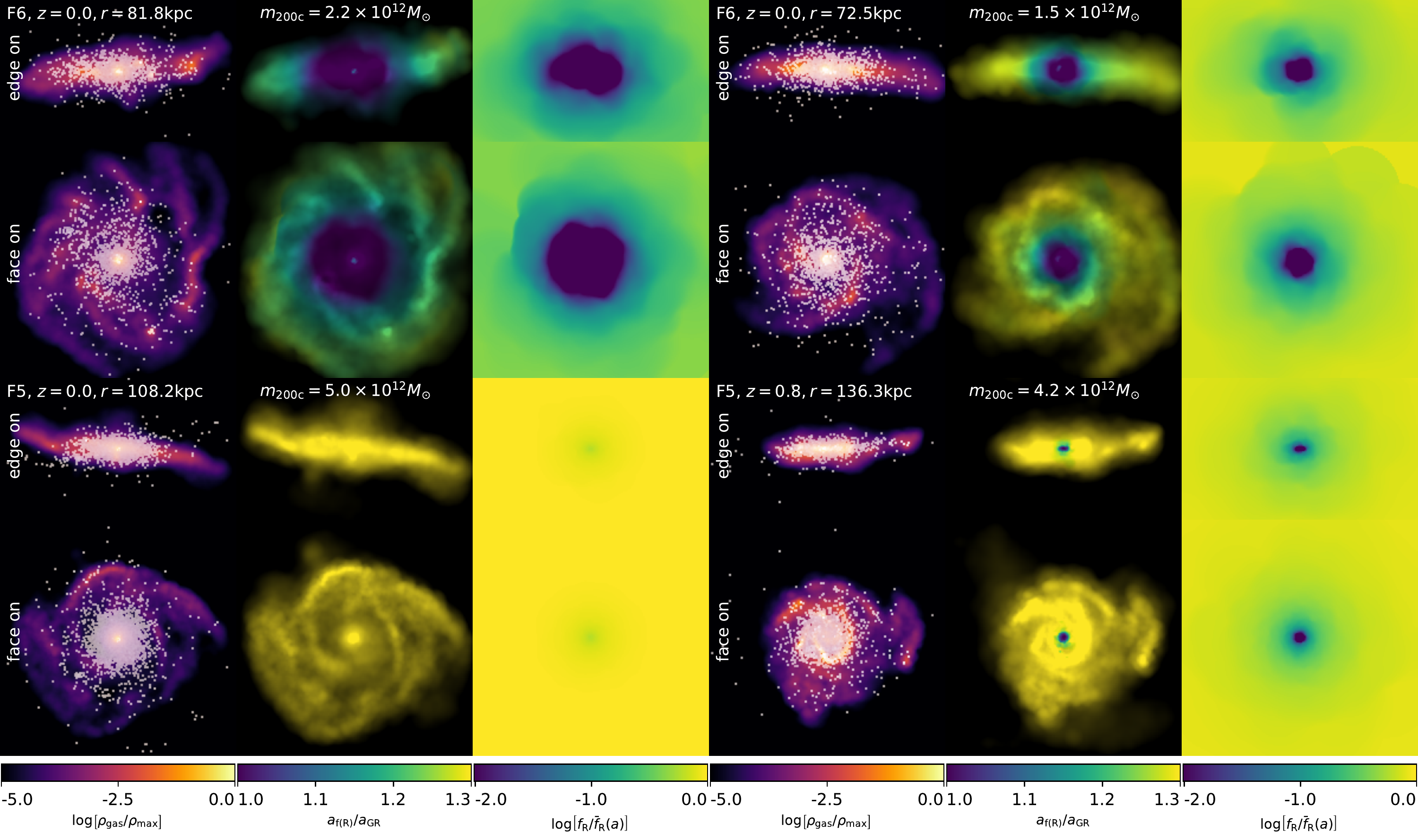}};
\draw[color = white, line width = 0.7mm] (0, 0.32592592592592595\textwidth) -- (\textwidth, 0.32592592592592595\textwidth);
\draw[color = white, line width = 0.7mm] (0.5\textwidth, 0.05925925925925926\textwidth) -- (0.5\textwidth, 0.593\textwidth);
\node[anchor = south west, inner sep = 0, red] at (0.47\textwidth, 0.33\textwidth) {\Huge{\bf a}};
\node[anchor = south west, inner sep = 0, red] at (0.97\textwidth, 0.33\textwidth) {\Huge{\bf b}};
\node[anchor = south west, inner sep = 0, red] at (0.47\textwidth, 0.0633\textwidth) {\Huge{\bf c}};
\node[anchor = south west, inner sep = 0, red] at (0.97\textwidth, 0.0633\textwidth) {\Huge{\bf d}};
\end{tikzpicture}
\caption{A selection of four galaxies from the F6 (top) and F5 (bottom) cosmology simulations at $z=0$ (top left, top right, and bottom left)  and $z=0.8$ (bottom right). For each galaxy, the top panels show an edge-on view where the angular momentum of the stars is pointing up; the lower panels show a face on view. The left panels show the gas column density for the individual galaxies combined with the stars within the objects. The middle panels show the gas density (brightness) color-coded with the modified gravity forces within a thin slice through the center of the galaxies. Yellow coloured gas cells experience an enhanced total force $F_{\rm tot} = 4/3 F_{\rm GR}$ while the dark blue regions do not experience a force enhancement due to modified gravity. The scalar field to back ground field ratio $f_{\rm R} / \bar{f}_{\rm R}(a)$ (see methods) within the same slice through the central region is shown in the right hand panels. Again, yellow regions are unscreened with $f_{\rm R} \approx \bar{f}_{\rm R}(a)$. The dark blue regions are fully screened $f_{\rm R} < 10^{-2} \bar{f}_{\rm R}(a)$. We quote model, redshift, size of the plotted region from the center of the object and the total mass within $r_{\rm 200 crit}$ in the plots for each of the galaxies. Note that rotationally supported disk galaxies can form in $f(R)$-gravity despite the complicated force morphology in the partially screened regime.} 
\label{fig:galaxies}
\end{figure*}
\begin{figure}
\centering
\includegraphics[width = \columnwidth]{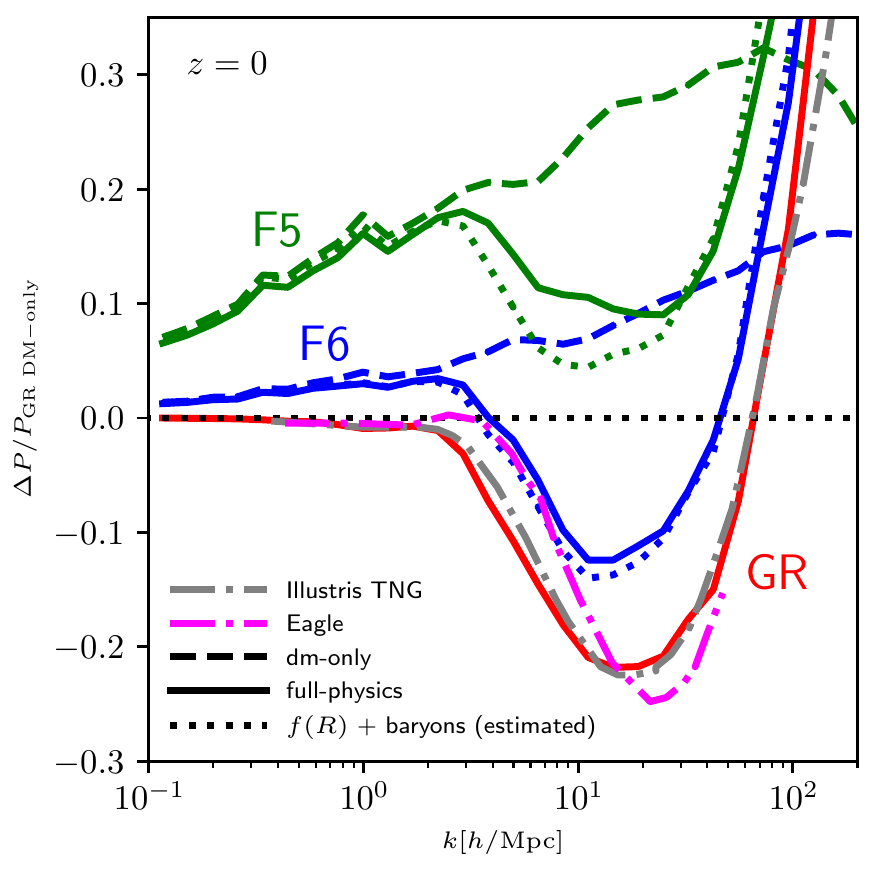}
\caption{The relative difference of the matter power spectra with respect to the $\Lambda$CDM cosmology DM-only simulation for a $\Lambda$CDM universe (\textit{red}), the F6 model (\textit{blue}) and F5 (\textit{green}). \textit{Dashed lines} show relative differences for the  DM-only $f(R)$-gravity runs, \textit{solid lines} for the full-physics simulations using the same colours as above. 
An estimate for the combined effect of baryonic feedback and modified gravity (displayed as \textit{dotted lines}) was obtained by adding the relative differences for  the $f(R)$-gravity DM-only simulations  to those from  the GR full physics run. 
The \textit{gray} and \textit{magenta dash-dotted lines} indicate the impact of feedback on the matter power spectrum in the Illustris TNG \protect\citep{springel2018} and Eagle \protect\citep{schaye2015} simulations, respectively.
The \textit{horizontal black dotted line} indicates equality.} 
\label{fig:power}
\end{figure}
\begin{figure*}
\centering

\begin{tikzpicture}
\node[anchor = south west, inner sep = 0]  at (0, 0) {\includegraphics[width = \textwidth]{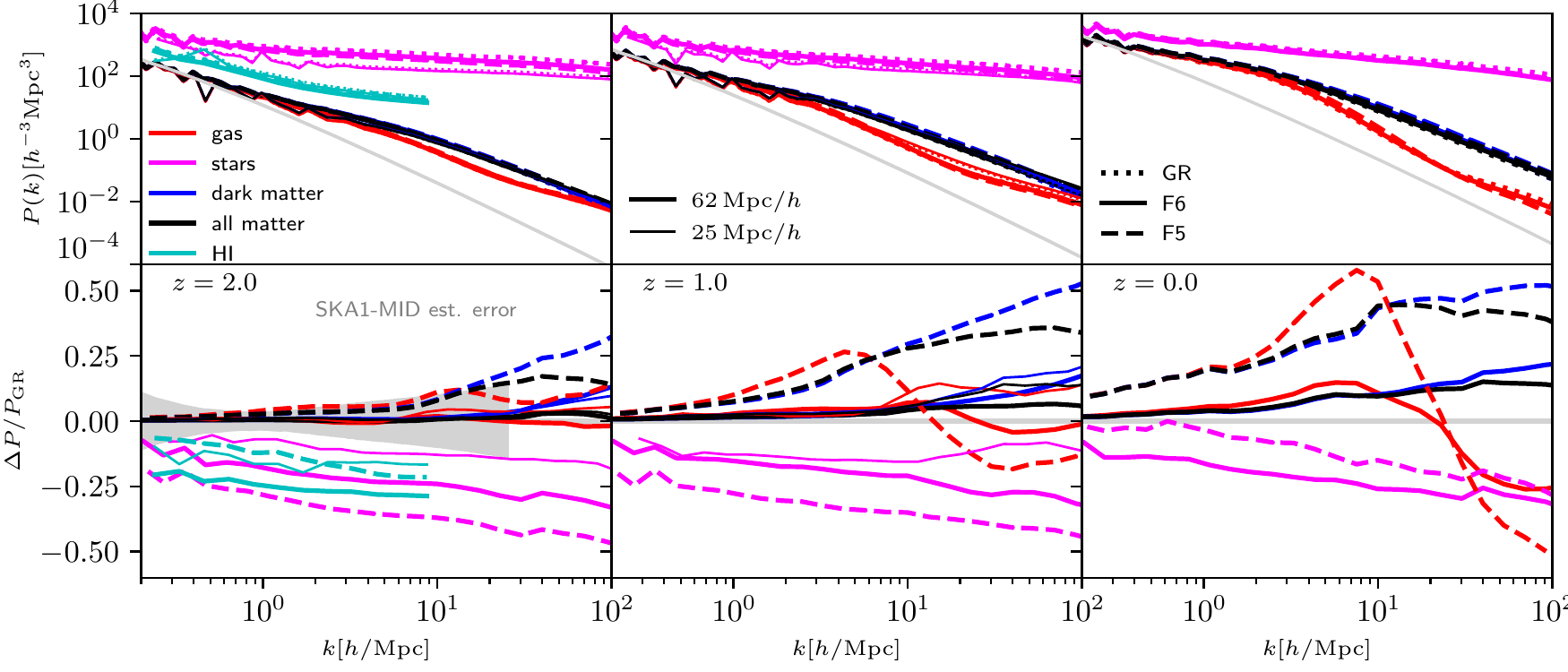}};
\node[anchor = south west, inner sep = 0, red] at (0.1\textwidth, 0.07\textwidth) {\Huge{\bf a}};
\node[anchor = south west, inner sep = 0, red] at (0.4\textwidth, 0.07\textwidth) {\Huge{\bf b}};
\node[anchor = south west, inner sep = 0, red] at (0.7\textwidth, 0.07\textwidth) {\Huge{\bf c}};
\end{tikzpicture}
\caption{The 3D matter power spectrum of the different matter components in our simulations for redshift $z=2$ (\textbf{a}), $z=1$ (\textbf{b}) and $z=0$ (\textbf{c}). GR results are shown as dotted lines, results from our F6  and F5 simulations as solid and dashed lines, respectively. Absolute values are shown in the upper panels while the relative differences are indicated in the lower panels for each redshift. Thick lines indicate results from the $62\, h^{-1}\textrm{Mpc}$ simulation box, thin lines results from the $25\, h^{-1}\textrm{Mpc}$ box (these are only shown for $z=2$ and $z=1$). The power spectrum of gas is displayed in red, stars in magenta, dark matter in blue and the total matter spectrum in black. For $z=2$ and $k < 10 \,h {\rm Mpc}^{-1}$, we show the neutral hydrogen power spectrum in cyan as well and compare it to predicted errors for SKA1-MID measurements of the HI power spectrum (gray shaded region; \protect\cite{ska2015}). In the upper panels, the grey lines indicate the linearly evolved initial power spectrum. The horizontal grey lines in the lower panels indicate equality.} 
\label{fig:power_components}
\end{figure*}
The physical nature of gravity and dark energy are central unsolved problems in modern physics. Einstein's General Relativity (GR) is generally accepted as the standard theory of gravity. 
GR has been empirically verified, to remarkably high precision, on small scales, but on cosmological scales the available tests are still not strongly constraining. 
Planned astronomical surveys will deliver data capable of distinguishing it from alternative theories of gravity. 
However, predictions for these observables require accurate simulations of the formation of cosmic structure in GR and alternatives, which must include the interactions between baryons and dark matter. 
Here, we present a set of cosmological hydrodynamical simulations of an attractive alternative to GR, the Hu \& Sawicki $f(R)$-gravity, which is representative of the large class of chameleon-type screening modified gravity models. 
These simulations are the first to include galaxy formation processes, such as feedback from supernovae and super-massive black holes, which can affect observables at similar levels to the modifications to GR. 
We find that it is possible to form rotationally supported disk galaxies in $f(R)$-gravity and that back-reaction effects on the matter power spectrum are small. 
We also find that the stellar and neutral hydrogen power spectra are suppressed in $f(R)$-gravity to a level that is detectable with future neutral hydrogen surveys.

The results presented in this work were derived from the SHYBONE (Simulating HYdrodynamics BeyONd Einstein) simulations, a set of full-physics hydrodynamical simulations employing the Illustris-TNG model in Hu-Sawicki $f(R)$-gravity \citep{buchdahl1970,husa2007}, which is one of the most widely-studied models of modified gravity. The theory introduces an extra scalar degree of freedom which mediates a fifth force between matter particles. This enhances standard gravity by up to $4/3$ in low-density environments, while dense regions are screened from such enhancements via the chameleon screening mechanism \citep{khoury2004}. This way, the model can pass the stringent constraints on gravity in the solar system \citep{will2014} and still leave detectable signatures on large scales, making it an excellent tool to explore how possible deviations from GR would be observable.

$f(R)$-gravity predicts identical propagation speeds of gravitational waves and photons, and so is compatible with the recent gravitational wave detection with optical counterparts \citep{lombriser2016,sakstein2017,ezquiaga2017,lombriser2017}. With appropriate choices of model parameters (see Methods), it can explain the late-time accelerated expansion of our Universe without an explicit cosmological constant \citep{husa2007} (Note that the value of $\Omega_\Lambda$ still enters the theory as a parameter). In addition to being theoretically well understood \cite{sofa2010, joyce2015}, $f(R)$-gravity has also been studied extensively in dark matter only (DM-only) \citep{schmidt2010, zhao2011, li2011, lombriser2013, puchwein2013, hellwing2013, zivick2015, cautun2018, mitchell2018, li2018} and non-radiative hydrodynamical cosmological simulations \citep{arnold2014, arnold2015, hammami2015, he2016}. These works were nevertheless unable to include a suitably calibrated baryonic feedback model at the same time, which is nontrivial and computationally expensive.

Such simulations with feedback have recently become possible thanks to a modified \textsc{arepo} \citep{springel2010} code, which employs a new and optimized method to solve the fully nonlinear $f(R)$-gravity equations in the quasi-static limit (see Methods). In this work, this more efficient code is combined with the Illustris-TNG galaxy formation model \citep{pillepich2018b, springel2018, genel2018, marinacci2018, nelson2018} which incorporates prescriptions for gas-hydrodynamics, magnetic fields, star and black hole formation, feedback from supernovae and AGN, gas heating and cooling processes, as well as galactic winds. This new code made the SHYBONE simulations possible, which are dedicated to in-depth studies of the interplay of baryonic physics and modified gravity for the first time. A detailed description of the simulations used in our analysis can be found in Methods.

\section{Results}
\label{sec:results}

\subsection{Galaxies in $f(R)$-gravity}

Figure \ref{fig:galaxies} shows a selection of galaxies from our full-physics simulations for the F6 (\textbf{a} and \textbf{b}) and F5 model (\textbf{c} and \textbf{d}) in the (total) mass range $1.5\times10^{12}<M_{\rm 200c}/M_\odot<5\times10^{12}$. Three of the objects are at $z=0$ (\textbf{a}-\textbf{c}), while the galaxy in Panel \textbf{d} is at $z = 0.8$ (only one disk galaxy is identified at $z=0$ for F5). The galaxies were selected to have a rotationally supported stellar disk, i.e., a rotational-to-total kinetic energy parameter $\kappa = \sum (j_z/r)^2 / \sum v^2>0.57$ (where the $z$-direction is defined by the total angular momentum of the stars in the galaxy)
\citep{ferrero2017}.

It is evident from the central sub-panels of Figure \ref{fig:galaxies} that both galaxies shown for the F6 model are partially screened. The gas in the inner regions is screened and experiences GR-like forces (dark blue regions; $a_{f(R)}/a_{\rm GR}\approx1$), while the gravitational force experienced by the gas in the outer regions is enhanced by $4/3$ (bright yellow regions in the gas). The size of the central screened region depends on the total mass of the galaxy and its host halo, and the results imply that objects similar to the Milky Way Galaxy are partially screened for F6. In the F5 model, in contrast, the chameleon screening is less efficient, and the objects shown in the bottom panels show either no (c) or very small (d) screened regions in the central panels. Gas and stars will thus experience fully enhanced gravity across the whole object, implying that the model is unlikely to pass local tests of gravity.

Interestingly, this complicated force morphology in F6 does not hinder galaxies from forming rotationally-supported disks within the gas and the stars. We find that there are more galaxies with $\kappa > 0.57$ in F6 compared to GR and significantly fewer in F5. The latter is an interesting observation that might be related to the stability of the disks in the unscreened regime or an enhanced galaxy merger rate. However, we caution that the number of disk galaxies is small in all of our simulations ($7,16,1$ objects at $z=0$ and $34,48,22$ objects at $z=0.8$ for GR, F6 and F5, in the large box, respectively and $56,64$ for GR and F6 at $z=1$ in the small box). The number counts of disk galaxies is therefore to be verified in future simulations.

\subsection{The matter power spectrum degeneracy}

It is well-known that $f(R)$-gravity can cause enhancements in the matter power spectrum while baryonic effects, predominantly AGN feedback, act in the opposite direction \cite{puchwein2013, arnold2018}. Previous simulation works were nevertheless unable to study both effects simultaneously and therefore could not model the possible back-reactions between them -- a challenge that the full-physics simulations in this paper aim to tackle.

To inspect this degeneracy, in Figure \ref{fig:power} we plot the fractional changes of the total matter power spectrum with respect to results from the GR DM-only simulation. The dashed lines show the enhancement predicted by the $f(R)$ DM-only simulations. Solid lines show the combined effects predicted by the full-physics $f(R)$ simulations (for GR they show the impact of baryons only). The dotted lines are produced by simply adding up the relative changes due to modified gravity, as predicted by $f(R)$ DM-only simulations, and due to AGN feedback, as predicted by the GR full-physics simulation.

As one can see from the plot, baryonic feedback suppresses power on scales $k\gtrsim2\,h {\rm Mpc}^{-1}$, with the suppression reaching $\sim20\%$ at scale $k\sim10\,h {\rm Mpc}^{-1}$. These results are in excellent agreement with previous works employing the same hydrodynamical model (\cite{springel2018}) and from the EAGLE simulation (\cite{hellwing2016}). Similar suppression at $k\gtrsim2\,h {\rm Mpc}^{-1}$ due to baryonic feedback can be seen for F6 and F5. 

It is obvious from the plot that the effect of baryons on the power spectrum for the F6 model is negligible for $ k < 2\,h {\rm Mpc}^{-1}$. The relative difference of the full-physics $f(R)$-gravity power spectrum is thus dominated by the modified gravity effect which increases power by a few percent at these scales. As for GR, the suppression due to baryons becomes active beyond $k = 2\,h {\rm Mpc}^{-1}$. Due to the $f(R)$-gravity effects, the minimum of the relative difference is nevertheless only $12 \%$ below the baseline. 
Interestingly, the blue dotted line in Figure \ref{fig:power} shows that the power spectrum of the full-physics F6 simulation can be predicted almost perfectly by an additive combination of the effects due to modified gravity and baryonic feedback. This suggests that there is negligible back-reaction between these effects for F6, and therefore computationally costly full-physics simulations for this model are not necessary as far as the matter power spectrum is concerned.
This simplification does not hold for the F5 model, where the back-reaction effect is strong enough that a full-physics simulation is necessary if percent-accuracy predictions of the matter power spectrum are needed even at $k\lesssim1\,h {\rm Mpc}^{-1}$. This will have important implications on the tests of this and similar models in surveys like Euclid.

The different strengths of interplay between modified gravity and AGN feedback in F6 and F5 can be explained by the chameleon screening mechanism. In the F6 model, the central regions of halos which are massive enough to carry AGN are screened, so that the flow of matter onto the central black hole (the AGN accretion) is not altered by modified gravity, resulting in similar AGN-feedback effects on the power spectrum as in GR. 
For  F5, on the other hand, many of the AGN-hosting halos have become completely unscreened.
The flow of matter towards central black holes therefore takes place in environments with by $4/3$ enhanced gravity, leading to modified accretion and feedback efficiency, and in turn a non-negligible back-reaction between the modified gravity and feedback effects in the matter power spectrum (for details on the different types of AGN feedback see \cite{pillepich2018}). 

\subsection{Power spectra of individual components}
\label{power_components}

The effect on the total matter power spectrum seen above is a combination of effects on the different matter components. In Figure \ref{fig:power_components} we display the power spectra of these individual components for $z = 2, 1$ and $0$ (Panel a, b and c, respectively). The absolute spectra, which we find to agree with the original Illustris-TNG simulation \citep{springel2018}, are displayed in the upper panels, the relative differences relative to GR in the lower panels. 

The components react in very different ways to modified gravity. The dark matter distribution is to first order only affected by gravity, its power spectrum is thus enhanced in a similar way as in DM-only simulations \citep[e.g.,][]{winther2015, arnold2018}. The dark matter and total matter power spectra are converged at $k\lesssim20\,h {\rm Mpc}^{-1}$ between the $25\, h^{-1}\textrm{Mpc}$ and $62\, h^{-1}\textrm{Mpc}$ boxes, suggesting that the results in Figure \ref{fig:power} can be trusted in this $k$-range.

The gas power spectrum is only mildly influenced by modified gravity at $z=2$, and the effects are stronger at lower $z$. In particular, the gas is more clustered in F6 and F5 than in GR at intermediate ($k<10\,h {\rm Mpc}^{-1}$) scales. 
These effects are caused by the interplay of a number of processes that we only attempt to explain to first order here: the enhancement in gas clustering on intermediate scales is mainly a result of higher gas densities in the unscreened parts of galaxies -- as Figure \ref{fig:galaxies} shows, this will be primarily the case in low-mass unscreened objects or the outer regions of intermediate-mass objects. Comparing the enhancements in the gas power spectrum for F6 at $z=1$ for the small and the large simulation boxes, we see that the results are consistent at few-percent level for $k<10\,h {\rm Mpc}^{-1}$.

Unlike dark matter and gas, the clustering of stars is more strongly effected by modified gravity at high redshift. At $z=2$, the stellar power spectrum from the $62\, h^{-1}\textrm{Mpc}$ box is suppressed by $\sim20$-$25\%$ in F6 and $\sim30$-$40\%$ in F5 on scales $k\sim1$-$10\,h {\rm Mpc}^{-1}$ (thick magenta lines). The predicted suppression for F6 from the $25\, h^{-1}\textrm{Mpc}$ simulation (thin magenta line) is weaker, which is $\sim10$-$15\%$ in the same $k$ range, indicating that the clustering of stars has not converged in the big box at this redshift. At $z=1$, the two simulation boxes agree to a few percent for $k<5\,h {\rm Mpc}^{-1}$.

A main reason for the suppression of the stellar power spectrum is the gas density in low-mass objects: as smaller halos are already unscreened at high redshift, the gas within these objects is denser and can cool more effectively, compared to the similar objects in GR. This leads to enhanced star formation inside small halos. We have checked that the fraction of $m = 10^{10} M_\odot$ halos carrying stars is in fact $60\%$ higher in F6 and $100\%$ higher in F5 compared to GR at $z=2$. These halos form from lower initial density peaks, which are intrinsically less clustered. This, together with the larger numbers of star-forming halos at high redshift in $f(R)$-gravity compared to GR, lead to a suppression of the stellar power spectrum in the former. As this process does not involve stellar dynamics, we expect it to be unaffected by potential self-screening of stars (which is not taken into account in our simulations).

\subsection{The neutral hydrogen power spectrum}

The enhanced gas cooling efficiency in low-mass objects in $f(R)$ gravity at early times also affects the distribution of neutral hydrogen (HI), a large fraction of which resides in small 
halos. We quantify this using the HI power spectrum, calculated following the procedure outlined in \cite{navarro2018}. 
The results from the $25\, h^{-1}\textrm{Mpc}$ boxes are plotted in Figure \ref{fig:power_components}(\textbf{a}), which shows that at $z=2$ the HI power spectrum on scales $k\sim0.02$-$10\,h {\rm Mpc}^{-1}$ is suppressed by $\sim15\%$ for F6 (thin cyan line). The $62\, h^{-1}\textrm{Mpc}$ simulation boxes have not yet reached convergence 
as they cannot resolve small halos ($m \lesssim 10^{10} M_\odot$) which host a significant fraction of the total HI. 
The small box, on the other hand, resolves halos down to $m\sim3\times10^8M_\odot$, which host the vast majority of HI \citep{navarro2018}. For both box sizes we checked that the relative suppression of the HI power spectrum in $f(R)$ gravity is almost identical to the suppression of the power spectrum of HI-hosting halos at $k\lesssim2\,h{\rm Mpc}^{-1}$, showing that it is more sensitive to the clustering of halos than to baryonic feedback.
Comparing our results to the expected uncertainty for an intensity mapping experiment similar to SKA1-MID (a future $21\,{\rm cm}$ intensity mapping survey, \citep{ska2015}), we find that with 1000 hours of observing time, SKA1-MID should be able to distinguish F6 and GR for $z=2$. This is different from most other cosmological tests of this model, which are usually at lower redshifts.

Also interestingly, at $z=2$ the HI power spectrum for F5 is suppressed less than F6, which is because at this redshift F6 has a larger enhancement of the number of small halos while in F5 such small halos have merged to form larger ones. We checked that the F5 HI power spectrum is suppressed more at even higher redshifts.

While both the HI fraction and star formation rate in low-mass objects are sensitive to (stellar) wind feedback, we expect the impact by $f(R)$-gravity to be relatively robust against changes in the feedback model. This is because such changes cannot be arbitrary as that would otherwise lead to tensions between the simulations and observational data for the stellar observables (see Figure B1 in \cite{pillepich2018} for a detailed discussion on how different wind feedback implementations affect the stellar properties of galaxies at $z=0$). It is also known \cite{altay2013} that details in wind feedback are important in the strong Damped Lyman-$\alpha$ regime where HI column density is $>10^{21.5}{\rm cm}^{-2}$, while we have checked explicitly that modified gravity effects are non-negligible only at column densities $\lesssim10^{21.2}{\rm cm}^{-2}$.


\subsection{Galaxy properties}
\label{gal_prop}

\begin{figure*}
\centering
\begin{tikzpicture}
\node[anchor = south west, inner sep = 0]  at (0, 0) {\includegraphics[width = 0.8\textwidth]{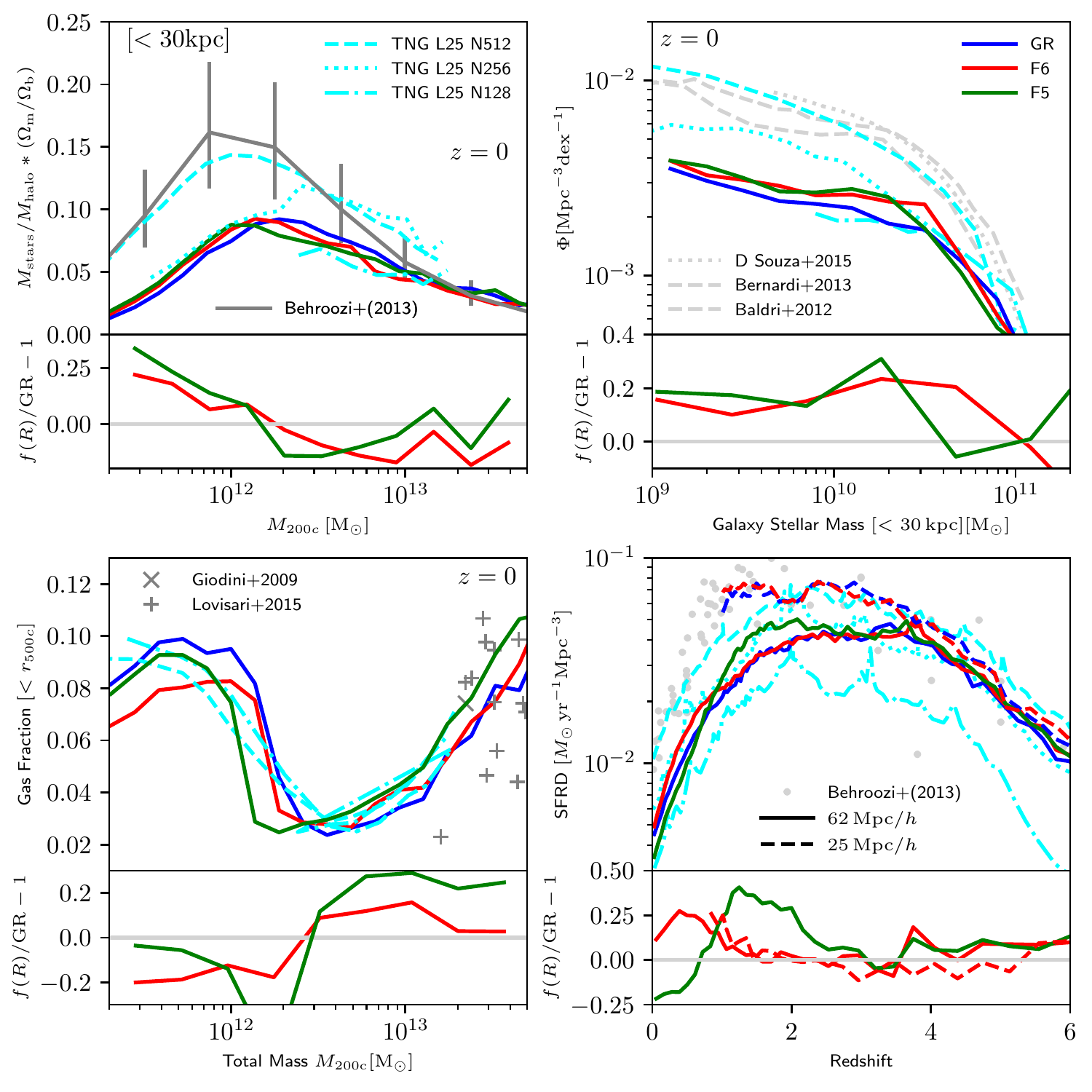}};
\node[anchor = south west, inner sep = 0, red] at (0.35\textwidth, 0.63\textwidth) {\Huge{\bf a}};
\node[anchor = south west, inner sep = 0, red] at (0.75\textwidth, 0.63\textwidth) {\Huge{\bf b}};
\node[anchor = south west, inner sep = 0, red] at (0.35\textwidth, 0.17\textwidth) {\Huge{\bf c}};
\node[anchor = south west, inner sep = 0, red] at (0.75\textwidth, 0.17\textwidth) {\Huge{\bf d}};
\end{tikzpicture}
\caption{The stellar and gaseous properties of galaxies. Results from the simulations presented in this work are shown as blue lines for GR, red lines for F6 and green lines for F5 in the upper section for each of the four panels. The lower sections show the relative difference between $f(R)$-gravity and GR.  The cyan lines show the results from \citep{pillepich2018} for the $25\, h^{-1}\textrm{Mpc}$ side-length Illustris-TNG test boxes with $2\times 128^3$ (dash-dotted lines), $2\times 256^3$  (dotted lines)  and $2\times 512^3$ (dashed lines) initial resolution elements. Observational constraints from \cite{behroozi2013, dsouza2015, bernardi2013, baldri2012, giodini2009, lovisari2015} are shown as grey lines and symbols. \textbf{a:} The galaxy stellar mass fraction within $30 {\rm kpc}$ from the halo centre as a function of halo mass in the $62\, h^{-1}\textrm{Mpc}$ box. \textbf{b:} The galaxy stellar mass function measured within $30 {\rm kpc}$ from the halo centre in the $62\, h^{-1}\textrm{Mpc}$ box. \textbf{c:} The galaxy gas fraction within $r_{500 {\rm c}}$ as a function of total halo mass in the $62\, h^{-1}\textrm{Mpc}$ box. \textbf{d:} The cosmic SFRD as a function of redshift for both the $25\, h^{-1}\textrm{Mpc}$ (dashed lines, for $z>1$) and the $62\, h^{-1}\textrm{Mpc}$ (solid lines) simulation boxes. } 
\label{fig:four}
\end{figure*}
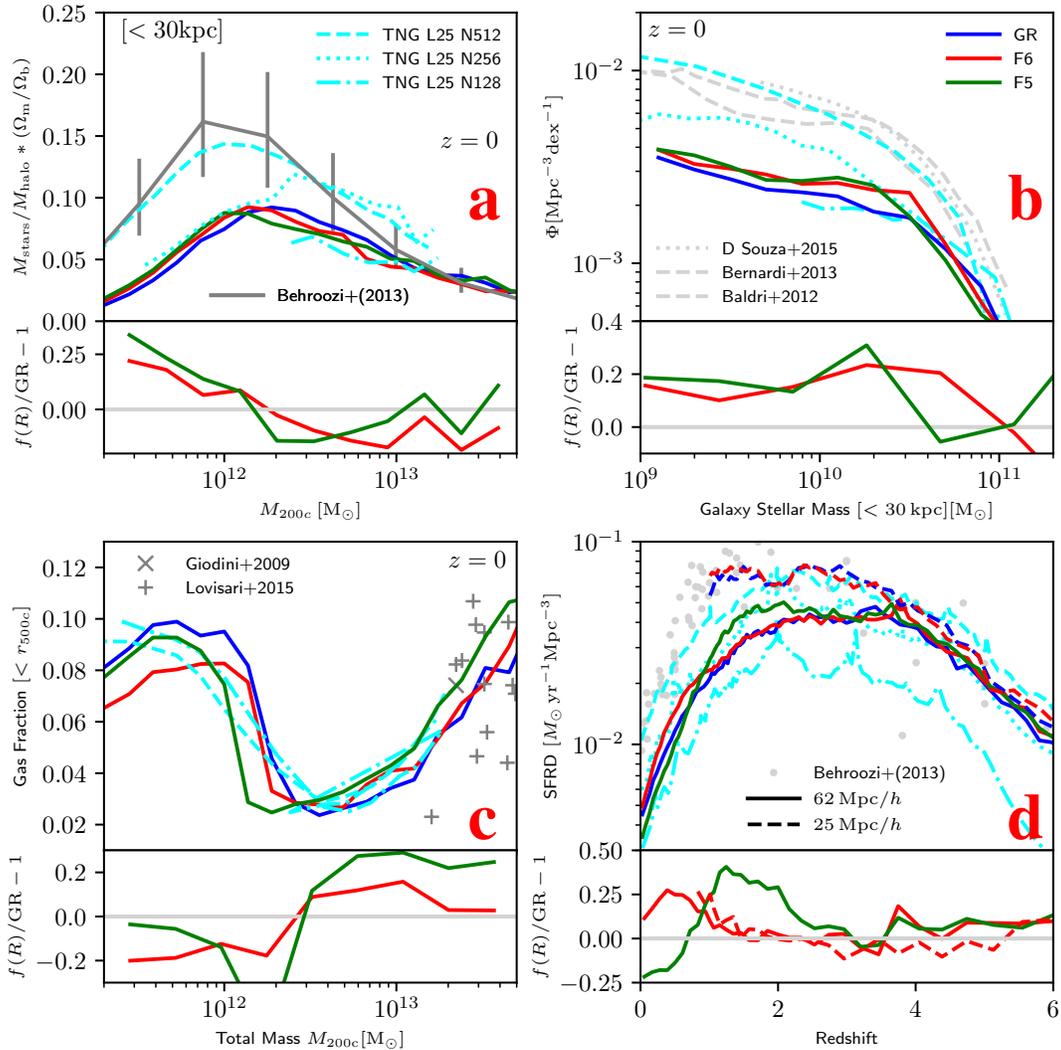

The SHYBONE simulations use the same hydrodynamical model for all three gravity theories. To check that a re-tuning of this model is not necessary for $f(R)$ gravity, in Figure \ref{fig:four} we plot a few stellar observables which were used in the tuning of the Illustris-TNG model. Panel \textbf{a} displays the stellar mass fraction (SMF) at the centres of galaxies, as a function of the host halo mass from the $62\, h^{-1}\textrm{Mpc}$ box. We find that $f(R)$-gravity introduces a $\lesssim25\%$ change with respect to GR for both F6 and F5, which is smaller than the $1\sigma$ errors of the observational data \citep{behroozi2013} shown for comparison. The absolute value of the SMF in our simulations does -- as expected -- not match the observations: The TNG model was tuned for a $25\, h^{-1}\textrm{Mpc}$ box with $2\times512^3$ initial resolution elements, while simulations at lower resolution show lower star formation rates \citep[][see also the cyan lines in Panel {\bf d}]{pillepich2018}. For comparison, we also show the SMF from TNG simulations at several resolutions (the cyan lines in Panel {\bf a}). The mass resolution of our $62\, h^{-1}\textrm{Mpc}$ box is close to that in the $25\, h^{-1}\textrm{Mpc}$ TNG box with $2 \times 128^3$ resolution elements, and so are the SMF results. Supplementary Material Figure 1 further confirms that the SMF from our $25\, h^{-1}\textrm{Mpc}$ box, which has the same resolution as the highest-resolution TNG box, agrees with observational data at $z=1$.

The galaxy stellar mass functions measured within $30\,\rm{kpc}$ from the halo centres are shown in panel \textbf{b}. Again, the results from the $62\, h^{-1}\textrm{Mpc}$ box fall in between the TNG runs with $2\times128^3$ and $2\times256^3$ resolution elements, due to the reduced star formation rate in low-resolution simulations. The relative differences between $f(R)$-gravity and GR are smaller than the uncertainties in the observational data (grey lines).

Panel \textbf{c} shows the galaxy gas fraction as a function of the host halo mass. Unlike the stellar properties, the gas fraction is not resolution dependent. The results from the simulations presented in this work are therefore consistent with the three $25\, h^{-1}\textrm{Mpc}$ TNG boxes. The differences between the gravity models are again much smaller than the spread in observational data from \cite{giodini2009,lovisari2015} at $M_{200c}\gtrsim10^{13}M_\odot$.

Finally, the cosmic star formation rate densities (SFRD) as a function of redshift are shown in panel \textbf{d} for the $62\, h^{-1}\textrm{Mpc}$ and $25\, h^{-1}\textrm{Mpc}$ (for $z>1$) boxes. The results from the three TNG boxes confirm that the SFRD is resolution dependent. Our $25\, h^{-1}\textrm{Mpc}$ simulations match the observational data of \cite{behroozi2013}, and are comparable to the TNG test-box with the same resolution. The $62\, h^{-1}\textrm{Mpc}$ boxes, as expected, show a lower SFRD which falls between the two TNG test-boxes with $2\times128^3$ and $2\times256^3$ resolution elements. The relative difference between F6 and GR is nevertheless converged.

The effects of $f(R)$-gravity on galactic properties are the consequence of the interplay of different physical processes. 
The star formation rate depends on gas density and temperature, with an enhanced gravity resulting in higher gas densities within galaxies and consequently a higher SFRD, as shown in Figure \ref{fig:four} {\bf d}. In particular, at high redshift the star formation rates in $f(R)$-gravity and GR are roughly the same because most objects are screened; towards lower redshifts, galaxies gradually become unscreened and the SFRD in the two $f(R)$ models is enhanced with respect to GR. Due to the less efficient chameleon screening, in F5 this process happens earlier than in F6. At even later times, the relative enhancement of the SFRD in $f(R)$-gravity starts to decrease and becomes negative for F5 at $z\lesssim0.7$. A reason for this is that the cold dense gas gets consumed faster in the $f(R)$ models, leaving less material for star formation at later times. 
The increased SFRD causes a higher stellar mass function (Panel \textbf{b}), as well as making a larger number of small halos star forming, explaining the weaker clustering of stars. 

In addition to the four stellar and gas properties shown in Figure \ref{fig:four}, we have also checked our simulation predictions of the galaxy size and black hole mass, which were used to calibrate the Illustris-TNG baryonic model too. In all these cases, the relative difference induced by $f(R)$-gravity does not lead to tensions with observational data, and so we conclude that a re-tuning of the baryonic model is not necessary.


\section{Discussion}
\label{sec:conclusions}

Studying galaxy formation and its application to test gravity models with full-physics hydrodynamical simulations used to be a formidable task, because of the large computational cost and the complexities in calibrating a realistic baryonic model. There were also doubts regarding whether meaningful cosmological signals can be cleanly extracted given the uncertainty in the subgrid physics model. In this work, we present a new suite of high-resolution simulations of $f(R)$-gravity which incorporate a large number of baryonic processes to reproduce a realistic population of galaxies as implemented in the Illustris-TNG model. While this model has been applied in $f(R)$ simulations without changes, we have checked various galaxy gas and stellar properties, and found our simulation predictions in $f(R)$-gravity to agree with observations and the TNG simulations of similar resolutions, so that a re-tuning is not necessary.

A key question addressed here is the degeneracy between baryonic feedback and modified gravity, which has important implications for physical quantities, such as the matter power spectrum, that are central to future galaxy surveys such as Euclid. We find that in weak models like F6 these two effects can be modelled separately and combined additively, with better-than-percent accuracy in the predicted matter power spectrum at $k\lesssim1\,h {\rm Mpc}^{-1}$, while for stronger models like F5 there is a back-reaction between them, necessitating costly full-physics simulations if sub-percent accuracy is desired. 

Despite the complicated screening profile in them, we find rotationally-supported disk galaxies can form even in partially screened halos. There is an indication that strong $f(R)$ models such as F5 produce fewer disk galaxies than GR, which can be caused by the more frequent galaxy mergers in the former; if this is confirmed by future larger simulations, it may offer a possible test that is insensitive to baryonic physics (which has little impact on the merger rate). More interestingly, we find that the enhanced number of small dark matter halos in $f(R)$-gravity can significantly alter the spatial distributions of cool dense gas, stars and neutral hydrogen, and as a result the latter can be used as tracers for the clustering of small halos, which may offer useful model tests with future observations such as 21cm intensity mapping.

As a final remark, we note that though the resolution of the SHYBONE simulations is unprecedented in the case of $f(R)$-gravity, their relatively small boxes inevitably place a limit on a statistical analysis due to cosmic variance. Further, while the results look promising, in some cases, e.g., neutral hydrogen, research is still scarce even in GR, and in this sense we hope that our results will serve as a reality check and a call for more detailed studies in both GR and alternative models.

\section{Methods}
\label{sec:methods}

The results presented in this work were derived from a suite of cosmological hydrodynamical simulations of $f(R)$-gravity which, for the first time, allow to study the interplay between detailed baryonic feedback and modified gravity in the same calculation. The simulations were carried out with the \textsc{arepo} \citep{springel2010} code, by combining the Illustris-TNG galaxy formation model \citep{pillepich2018b, springel2018, genel2018, marinacci2018, nelson2018} and a new, much more efficient, modified gravity solver in the code. Below we give an overview of the TNG model and $f(R)$-gravity, and introduce the new modified gravity solver. Further details about the code verification tests are included in the Supplementary Material. 

The SHYBONE suite mainly consist of six simulations carried out for $\Lambda$CDM, F6 and F5 universes using the same initial condition. For each model, we carried out a full-physics hydrodynamical simulation and an associated DM-only run. These simulations initially contain $512^3$  dark matter particles and the same number of gas cells (for the hydrodynamical simulations only) in a $62\, h^{-1}\textrm{Mpc}$-a-side box, with mass resolutions $m_{\rm DM}=1.28\times10^8h^{-1}M_\odot$ and $m_{\rm gas}\approx 2.5\times10^7h^{-1}M_\odot$ (the mass within the gas cells can vary by a factor of $2$). 
For convergence test, two additional full-physics simulations (for F6 and GR) employing the same number of resolution elements in a $25\, h^{-1}\textrm{Mpc}$ box, with $m_{\rm DM}=8.39\times10^6h^{-1}M_\odot$ and $m_{\rm gas}\approx 1.6\times10^6h^{-1}M_\odot$,  were run until $z=1$.
All simulation use the same starting redshift, $z=127$. The softening lengths for DM particles and stars are $0.5$ and $1.25\,h^{-1}{\rm kpc}$ for the small and large boxes, respectively. The total run time is $\sim2.5$ million core hours, and all simulations use Planck 2016 \citep{planck2016} cosmological parameters $\sigma_8 = 0.8159$, $\Omega_m = 0.3089$, $\Omega_B = 0.0486$, $\Omega_\Lambda = 0.6911$, $ h = 0.6774$ and $n_{\rm s} = 0.9667$.

Unless stated otherwise, we measure total masses of halos in terms of $m_{200\, {\rm c}}$ which is the mass within a sphere enclosing $200$ times the critical density of the Universe.

\subsection{Baryonic physics and feedback}

The Illustris TNG model, a successor of the Illustris galaxy formation model \citep{vogelsberger2014}, incorporates prescriptions for a variety of physical processes necessary to reproduce a realistic galaxy population in cosmological simulations. The TNG model consists of a magneto-hydrodynamics solver on a moving mesh, a recipe for stellar formation, evolution and feedback, a prescription for black hole growth and AGN feedback \citep{weinberger2017}, as well as an algorithm for the chemical enrichment and cooling of the gas. It has been tuned to reproduce observational results on the stellar mass function, cosmic star formation rate density (SFRD), the relation between black hole mass and galaxy stellar mass, galaxy gas fraction, stellar mass fraction and the size of galaxies, in $\Lambda$CDM simulations \citep{pillepich2018}. For a more detailed overview of the TNG model, see \cite{pillepich2018b, vogelsberger2014b}.

For the simulations carried out for this project, we applied the same TNG model to all gravity theories considered. We are primarily interested in the question how astrophysical and cosmological observables in baryonic simulations are affected by modifications to gravity given a fixed model of galaxy formation. We explicitly checked that the changes to the observational gas and stellar properties against which the TNG model has been tuned are small compared to the uncertainties in the observations, so that a re-tuning is not necessary.

\subsection{$f(R)$-gravity}
\label{sec:gravity}

$f(R)$-gravity \citep{buchdahl1970} is an extension of Einstein's general relativity (GR) including an additional, scalar degree of freedom which induces an enhancement to the gravitational force in low-density environments. In dense regions like the Solar system, the model employs the chameleon screening mechanism \citep{khoury2004} to screen the modifications to gravity and recover GR-like behaviour.

The theory is constructed by adding a scalar function $f(R)$ of the Ricci scalar $R$ to the Einstein-Hilbert action of GR:

\begin{align}
S=\int {\rm d}^4x\, \sqrt{-g} \left[ \frac{R+f(R)}{16\pi G} +\mathcal{L}_m \right],\label{action}
\end{align}

where $G$ is the gravitational constant, $g$ is the determinant of the metric $g_{\mu\nu}$ and $\mathcal{L}_m$ is the matter Lagrangian density. Varying this action with respect to the metric leads to the Modified Einstein equations:

\begin{align} 
G_{\mu\nu} + f_{\rm R} R_{\mu\nu}-\left( \frac{f}{2}-\Box f_{\rm R}\right) g_{\mu\nu} - \nabla_\mu \nabla_\nu f_{\rm R} = 8\pi G T_{\mu\nu} \label{Eequn},  
\end{align} 

where $G_{\mu\nu}$ and $R_{\mu\nu}$ denote the Einstein and Ricci tensor, respectively, $T_{\mu\nu}$ the energy momentum tensor and $\Box \equiv \nabla_\nu \nabla^\nu$, where Einstein summation is used.
The derivative of the scalar function, $f_{\rm R} \equiv \d f(R)/ \d R$, becomes a new degree of freedom whose dynamics is determined by the trace of the modified Einstein equation.

In cosmological simulations one often works in the weak-field and quasi-static limit (see \cite{sawicki2015} for a discussion of its validity for $f(R)$-gravity), where Eq.~(\ref{Eequn}) simplifies considerably: similar to the Newtonian limit of GR, the gravitational potential $\Phi$ is given by a modified Poisson equation,

\begin{align}
 \nabla^2 \Phi = \frac{16\pi G}{3}\delta\rho - \frac{1}{6} \delta R,\label{poisson}
\end{align}

where $\delta\rho$ is the matter density perturbation, and $\delta R = R-\bar{R}$ is the perturbation to the Ricci scalar $R(f_R)$, with $f_R$ satisfying the following differential equation,

\begin{align}
 \nabla^2 f_{\rm R} =  \frac{1}{3}\left(\delta R -8\pi G\delta\rho\right). \label{fRequn}
\end{align}

Throughout this work we adopt the model proposed by \cite{husa2007}:

\begin{align}
 f(R) = -m^2\frac{c_1\left(\frac{R}{m^2}\right)^n}{c_2\left(\frac{R}{m^2}\right)^n +1},\label{fr}
\end{align}

$c$ denotes the speed of light here and $m$ is a mass scale of the model given by $m^2 \equiv \Omega_m H_0^2$, with $H_0$ being the Hubble constant, and $c_1$, $c_2$ are model parameters. We choose $n=1$ in this work (This is somewhat arbitrary but the most convenient choice for the computations and widely used in simulations of $f(R)$-gravity). 
In addition to its relatively simple functional form, this model has two further advantages. First, it allows the so-called chameleon screening mechanism to suppress the modifications to GR in high-density environments, which is necessary to pass the very tight constraints on gravity within the Solar system \citep{will2014}.
Second, the model features a cosmic expansion history which is very close to that of a $\Lambda$CDM universe if one chooses \citep{husa2007}

\begin{align}
\frac{c_1}{c2} = 6 \frac{\Omega_\Lambda}{\Omega_m} && \text{and} && \frac{c_2\, R}{m^2} \gg 1.\label{conditions}
\end{align}

Given the functional form in Eq.~\eqref{fr}, under the conditions of Eq.~\eqref{conditions} the scalar field $f_R$ can be approximated as

\begin{align}
f_{\rm R} \equiv \frac{\d f(R)}{\d R} \approx -\frac{c_1}{c_2^2}\left(\frac{m^2}{R}\right)^{2},\label{fR}
\end{align}

and consequently $R$ can be written as a function of $f_R$. Instead of $m,c_1,c_2$, the theory can be more conveniently described by the parameters $\Omega_m,\Omega_\Lambda$ and $\bar{f}_{\rm R0}$, with $\bar{f}_{\rm R0}$ having the physical meaning of the present-day value of the background scalar field. With these parameters specified, the scalar field at arbitrary time can be expressed as

\begin{align}
\bar{f}_{\rm R}(a)  = \bar{f}_{R0} \left[ \frac{\bar{R}_0}{\bar{R}(a)} \right]^2,\label{fRa}
\end{align}

where $R_0$ is the value of the Ricci scalar today and $R(a)$ is its value at scale factor $a$, 

\begin{align}
\bar{R} = 3m^2\left[ a^{-3} + 4\frac{\Omega_\Lambda}{\Omega_m} \right]. \label{R_bar}
\end{align}

The threshold for the onset of chameleon screening is set by $\bar{f}_{\rm R0}$. 
Current astrophysical constraints on the theory limit the background scalar field to values smaller than $|\bar{f}_{\rm R0}| = 10^{-6}$ (F6); cosmological constraints are weaker \citep{terukina2014} (\cite{he2018} find that the motions of galaxies in F6 predicted by DM-only simulations and sub-halo abundance matching (SHAM) are in tension with SDSS data. 
The authors of this paper nevertheless assume, that the same SHAM model can be applied to GR and f(R)-gravity if the mass of halos in the DM-only simulation, which is used as a parameter of the SHAM model in GR, is replaced by the dynamical mass for $f(R)$-gravity. The dynamical or effective mass is the mass a massive test particle would feel under the assumption of standard gravity. This mass measure accounts for the by $4/3$ enhanced gravitational forces in unscreened objects and is the same as the true mass in screened objects. This assumption is to be validated against full-physics MG simulations to asses the robustness of the constraint). We also consider a model with $|\bar{f}_{\rm R0}| = 10^{-5}$ (F5) within this work as a toy model which features more prominent effects due to modified gravity in the simulations but is in tension with observations \citep{terukina2014}.



\subsection{A new modified gravity solver in {\sc arepo}}
\label{sec:solver}

\textsc{arepo} \citep{springel2010} is a massively-parallel, highly-optimized, cosmological simulation code. It employs a second-order Riemann solver using a moving Voronoi mesh to solve the Euler equations of hydrodynamics. The code also features state-of-the-art prescriptions of star formation, feedback from supernovae and active galactic nuclei, and magneto-hydrodynamics \citep{vogelsberger2014, weinberger2017, pillepich2018}. 

The original gravity solver of \textsc{arepo} is based on the solver in \textsc{p-gadget3}, which is itself based on \textsc{gadget2} \citep{springel2005c}. The solver employs an  oct-tree algorithm to obtain the short-range gravitational forces while the long-range forces are calculated using a particle-mesh algorithm to reduce the communication load when running in parallel. \textsc{arepo}'s gravity solver has recently been improved and extended in several ways but for the implementation of the modified gravity solver we (currently) stick to the original version. 

Because of the similarity between the gravity algorithms in \textsc{p-gadget3} and \textsc{arepo}, we use the modified gravity solver in \textsc{mg-gadget} (\cite{puchwein2013},  also based on \textsc{p-gadget3}) as the starting point for the implementation of the $f(R)$ gravity solver in \textsc{arepo}. As the first step, we ported the solver from \textsc{gadget} to \textsc{arepo}, carrying out the necessary changes to adapt it to the code structure of \textsc{arepo} and optimizing the data structures of the solver to reduce its memory footprint. We will refer to this implementation as the `old' method in the following. In order to improve the convergence behaviour of the multigrid solver, we implemented a `new' method to solve for the scalar field based on the $f(R)$ gravity solver in \textsc{ecosmog} \citep{li2012} developed by \cite{bose2017}.

In order to ensure that the new $f(R)$ solver implemented in the code gives the correct result, we performed extensive comparison tests between the old and the new methods, as well as convergence tests for the MG-timestep criterion and the residual threshold on the individual grid levels. A selection of these tests are presented in the Supplementary Material.

\subsubsection{AMR-grid construction and mass assignment}

In order to solve for the gravitational forces one has to solve Eq.~(\ref{fRequn}) first before the scalar field can be used to calculate the gravitational potential through Eq.~(\ref{poisson}). Due to the chameleon screening mechanism, which causes the scalar field value to drop by many orders of magnitude in dense environments with respect to the background value $\bar{f}_{\rm R}(a)$, Eq.~(\ref{fRequn}) is highly nonlinear. 
The commonly-adopted method in cosmological simulation codes is therefore to solve the equation iteratively on a grid \citep{li2012, puchwein2013, llinares2014}. All modified gravity codes optimized for cosmological simulations employ an adaptively refining mesh (AMR-grid) for this purpose to allow for high spatial resolution in high-density regions while keeping the total computational cost down to a realistic level \citep{winther2015}.

In \textsc{arepo} we use an AMR-grid to solve for $f_{\rm R}$ as well. This is constructed from the gravitational oct-tree of the standard gravity solver in the same way as in \textsc{mg-gadget}. Up to a maximum level, each tree node represents a mesh cell.  In order to avoid a mesh that is too sparse on the finer AMR levels, all eight daughter tree nodes are created whenever a father node contains more than one simulation particle (this differs from the standard tree construction in \textsc{arepo}; for more details see \cite{puchwein2013}). Because the tree is constructed from scratch for every time step in \textsc{arepo}, there is no need to take care of tree-updates for this implementation. It is nevertheless necessary to construct the tree and solve for $f_{\rm R}$ before computing the long range gravity forces on the PM-grid as opposed to the standard gravity implementation in the code, where the tree is stored during the short-range force calculation only \citep{springel2010}. 

As in \textsc{mg-gadget}, the particle masses are assigned to the mesh using a cloud-in-cell (CIC) assignment algorithm allowing to calculate the density contrast field for each mesh cell. The following step, to actually solve for the scalar field $f_{\rm R}$, is where the old and new methods differ.

\subsubsection{Scalar field solver}

In the old method, Eq.~(\ref{fRequn}) is rewritten in the following form by defining a new variable $u = \ln\left(f_{\rm R}/\bar{f}_{\rm R}(a)\right)$,

\begin{align}
\nabla^2 {\rm e}^u = - \frac{1}{3c^2\bar{f}_{R}(a)} \left[ \bar{R}(a) \left( 1 - {\rm e}^{-\frac{u}{2}} \right) + 8 \pi G \delta\rho \right],\label{exp_method}
\end{align}

which is solved using the Newton-Raphson relaxation scheme with red black sweeps and multigrid acceleration (see \citep{puchwein2013} for more detail).
This way unphysical positive values of $f_{\rm R}$ which might occur due to numerical errors during the iterations can be avoided.


The new method makes use of a different re-parametrisation of $f_{\rm R}$. Following \cite{bose2017}, we define 

\begin{align}
u \equiv \sqrt{\frac{f_{\rm R}}{\bar{f}_{\rm R}(a)}},\label{u}
\end{align}

which prevents positive values for $f_{\rm R}$ as well. 
This leads to a different form of the scalar field equation in terms of $u$,

\begin{align}
\nabla^2 \left(u^2\right) = \frac{1}{3} \left[ \frac{\bar{R}(a)}{\bar{f}_{\rm R}(a)} \left(\frac{1}{u} -1 \right) - \frac{8 \pi G}{\bar{f}_{\rm R}(a)} \delta\rho  \right]. \label{laplace_usq}
\end{align}

This equation is then discretised on the AMR-grid constructed before. On a mesh of cell side-length $l$, the Laplace operator becomes

\begin{eqnarray}
&& \nabla^2 \left(u_{\rm i,j,k}^2\right)\nonumber\\ 
&=& \frac{1}{l^2}\left[ \left(u^2_{\rm i+1,j,k} + u^2_{\rm i-1,j,k} +  u^2_{\rm i,j+1,k} + \dots \right) - 6 u^2_{\rm i,j,k}\right]\nonumber\\
&\equiv& \frac{1}{l^2}\left[ L_{\rm i,j,k}(u^2) - 6 u^2_{\rm i,j,k} \right], \label{laplace_op}
\end{eqnarray}

where we defined the operator $L$. Further defining 

\begin{align}
\alpha \equiv \frac{\bar{R}(a)}{18 \bar{f}_{\rm R}(a)} && \beta \equiv \frac{8 \pi G}{18 \bar{f}_{\rm R}(a)},\label{alpha_beta}
\end{align}

and 

\begin{align}
p &\equiv -\frac{L_{\rm i,j,k}(u^2)}{6} - \alpha l^2 - \beta l^2 \delta\rho, \label{p} \\
q &\equiv - \alpha l^2,\label{q} 
\end{align}

the discretised version of Eq.~(\ref{laplace_usq}) becomes \cite{bose2017}

\begin{align}
u_{\rm i,j,k}^3 + p\, u_{\rm i,j,k} +q = 0. \label{cubic}
\end{align}

This cubic equation has a few advantages over the discretised equation in the old method. First, the result of the cubic equation can be found analytically for each mesh cell in each iteration step. The Newton-Raphson numerical update rule of the old method is thus avoided, leading to improved convergence speed and numerical stability of the solver. Second, the large number of expensive exponential and logarithm operations for each iteration step and mesh cell associated with the definition of $u$ are avoided. Finally, the new method does not introduce extra nonlinearities -- through the exponentials in Eq.~(\ref{exp_method}) -- in screened regions which slowed down the convergence of the solver in the old method, particularly at high redshift. 

Applying Cardano's method in order to solve Eq.~(\ref{cubic}), one has to distinguish the three branches of solutions depending on the values of $p$ and $q$. In order to be consistent with \citep{bose2017}, we define $\Delta_0 \equiv -3p$ and $\Delta_1 \equiv 27q$. Independent of the density within a given cell and the field value of its neighbours, $q < 0$ and thus $\Delta_1 < 0$ in all cases. The behaviour of the solution depends on the sign of the discriminant $D$:

\begin{align}
D \equiv \left( \frac{q}{2}\right)^2 + \left( \frac{p}{3}\right)^3.\label{discri}
\end{align}

\textbf{A:} For the case $D>0$, Eq.~(\ref{cubic}) takes only one real solution which is given by

\begin{align}
&& \tilde{u}_{\rm i,j,k} &= -\frac{1}{3}\left[C + \frac{\Delta_0}{C}\right],\nonumber\\
\text{where} && C &\equiv \left\{ \frac{1}{2} \left[ \Delta_1 + \left( \Delta_1^2 - 4 \Delta_0^3 \right)^{\frac{1}{2}} \right] \right\}^{\frac{1}{3}}.\label{Dgz}
\end{align}

We have used $\tilde{u}_{\rm i,j,k}$ to stress that the solution obtained is an approximation to the true solution $u_{\rm i,j,k}$ because the quantity $L_{\rm i,j,k}$ in $p$, cf.~Eq.~(\ref{p}), depends on the values $u_{\rm i\pm1,j\pm1,k\pm1}$ from the the neighbouring cells.
Note that this branch of the solution includes the case $p = 0$, for which Eq.~(\ref{Dgz}) simplifies to
$\tilde{u}_{\rm i,j,k}= \sqrt[3]{-q}$.

\textbf{B:} For $D < 1$, Eq.~(\ref{cubic}) admits three real solutions:

\begin{align}
&& 
\tilde{u}_{\rm i,j,k} &= - \frac{2}{3}\sqrt{\Delta_0} \cos\left( \frac{\Theta}{3} + \frac{2}{3} j \pi \right),\\ \label{Dsz}
\text{where} &&  \Theta &= \cos^{-1}\left(\frac{\Delta_1}{2\Delta_0^\frac{2}{3}}\right),
\end{align}

and $j \in \{ 0,1,2 \}$. The cases $j = 0$ and $j = 2$ lead to $\tilde{u}_{\rm i,j,k}<0$ which is in contradiction with our definition of $u$. The only physical solution thus is $j=1$.

After the cubic equation is solved for a cell, the value of $u$ therein is replaced by the above solution. In order to allow for efficient parallelisation on distributed-memory compute clusters and to allow for a faster convergence, we carry out the cell updates using a red-black-sweep pattern. We start from an initial guess of $u=1$ for the first time step of the simulation and use the result of the previous time step as an initial guess for all subsequent steps. 

At the end of each iteration, an approximate error for $u_{\rm i,j,k}^2$ ($\propto f_{\rm R}$) is calculated as 

\begin{equation}
e_{\rm i,j,k} = \frac{r_{\rm i,j,k}}{\bar{\rho}(a) l^2 \beta},\label{error}
\end{equation}

where

\begin{equation}
r_{\rm i,j,k} = \tilde{u}_{\rm i,j,k}^3 + p\, \tilde{u}_{\rm i,j,k} +q,\label{residual}
\end{equation}

is the residual for a given approximate solution $\tilde{u}_{\rm i,j,k}$.
We consider the solution on a given AMR-level to be converged and stop the iterations once 

\begin{align}
\max(e_{\rm i,j,k}) < 10^{-2}.\label{convergence}
\end{align}

This criterion translates to an (approximate) maximum error of $1\%$ on $f_{\rm R}/\bar{f}_{\rm R}(a)$ on a given grid level. We note that the resulting fifth force errors are smaller than the random standard gravity errors originating from the tree walk.

\subsubsection{Multigrid acceleration}

The relaxation method used above successively reduces the error of the initial guess by iterations. It is well known that the decay of long wave-length modes of the error can be dramatically sped up by multigrid methods. For both the old and the new method we therefore use multigrid acceleration arranged in terms of V-cycles to speed up the convergence of the scalar field solver. Our V-cycle implementation for the old method is identical to \textsc{mg-gadget}, and so we refer the reader to \cite{puchwein2013} for further details.

For the new method, we use the standard nonlinear full approximation storage (FAS) algorithm \cite{numerical}. Eq.~(\ref{cubic}) is rewritten in terms of a non-linear operator

\begin{align}
 \mathcal{L}_l(u_l) = 0, && \text{where} && \mathcal{L}_l(u_l) \equiv u_l^3 + p_l u_l + q_l.\label{nonlin}
\end{align}

The subscript $_l$ denotes quantities and operators defined on the `fine' grid level with cells of side-length $l$ here, we will use $_L$ for quantities and operators defined on the `coarse' level with cells of side-length $L = 2l$. Eq.~(\ref{residual}) can be written as 

\begin{align}
 \mathcal{L}_l(\tilde{u}_l) = r_l,\label{fine}
\end{align}

for some approximate solution $\tilde{u}_l$. After a few pre-smoothing iterations to ensure that the short-wavelength modes in $\tilde{u}_l$ have been reduced sufficiently, both sides of Eq.~(\ref{fine}) are mapped to the coarse level using the restriction operator

\begin{align}
\mathcal{R}\tilde{u}_l \equiv \sum\limits_{i=0}^8 \frac{\tilde{u}_l^i}{8},\label{restriction}
\end{align}

where the summation is over the 8 daughter cells of the coarse cell. This leads to the following equation on the coarse level

\begin{align}
\mathcal{L}_L(u_L) = \mathcal{L}_L(\mathcal{R}\tilde{u}_l)  - \mathcal{R} r_l, \label{coarse}
\end{align}

which can be solved either through iterations or further mapping to even coarser levels, to obtain an approximate coarse-level solution $\tilde{u}_L$. Using the prolongation operator $\mathcal{P}\tilde{u}_L = \tilde{u}_L$ the coarse-level solution can then be used to correct the fine-level solution as

\begin{align}
\tilde{u}^{\rm new}_l = \tilde{u}_l + \mathcal{P} (\tilde{u}_L - \mathcal{R}\tilde{u}_l). \label{correction}
\end{align}

This correction is applied in our multigrid implementation as follows. If no initial guess for $u$ is available, the solver starts to solve Eq.~(\ref{cubic}) iteratively on level 3 (the top-level node covering the whole simulation volume is level 1), and maps down this solution as the initial guess for the next finer level. This process is continued recursively until the finest level that covers the whole simulation volume (max-level-full-tree, MLFT) is reached. If an initial guess is available, the solver starts on MLFT directly. 

All the finer levels (including MLFT) are then solved by V-cycles using corrections from the two next coarser grid levels. The solver would, for example, repeat the cycles 7-6-5-6-7 on level 7 until convergence is reached and then continue on level 8 with 8-7-6-7-8 V-cycles using the solution of level 7 as the initial guess on level 8. 
During the V-cycles, the solver carries out 4 pre- and 4 post-smoothing iterations before mapping to and after mapping back from the coarser level, to ensure that the short-wavelength modes of the error of $\tilde{u}$ have sufficiently decayed. If the solver determines that the V-cycles do not lead to convergence on a given level, that level will be solved by iterations only. 
On the finest level which the mesh refinements extend to, the solution is obtained by iterations only for stability reasons. 

\subsubsection{Force calculation}

Both the old and new methods adopt the effective mass approach to calculate the total gravity force as in \textsc{mg-gadget}, making use of the fact that the modified Poisson equation, (\ref{poisson}), can be rewritten in terms of an effective density 

\begin{align}
\nabla^2 \Phi = 4\pi G (\delta\rho + \delta\rho_{\rm eff}), \label{Peff}
\end{align}

where 

\begin{align}
\delta\rho_{\rm eff} = \frac{\delta\rho}{3} - \frac{\delta R}{24\pi G}. \label{rho_eff}
\end{align}

The advantage of this approach is that it allows to calculate the gravitational accelerations from Eq.(\ref{Peff}) using the standard oct-tree -- particle-mesh gravity solver readily implemented in \textsc{arepo}.

\subsubsection{Local time stepping}

To keep the number of computationally expensive modified gravity force calculation low, we adopt the local time-stepping scheme that was first implemented in \textsc{mg-gadget} (for more details see \cite{arnold2016}). This scheme takes advantage of the fact that the regions which require the smallest time steps, i.e., the regions with the highest accelerations, lie primarily within massive halos and are thus screened by the chameleon mechanism. For reasonable values of $f_{\rm R}$, the maximum modified gravity acceleration in the simulation box at a given time is orders of magnitude smaller than the standard gravity acceleration. Decoupling the modified gravity time step from the standard gravity and hydro time steps can therefore significantly reduce the computational cost of the simulations without loss of accuracy. 

In practice, we decouple these time steps using a local time-stepping scheme already implemented in \textsc{arepo}. The modified gravity time step is coupled to the long-range gravity PM time step global, while the short range gravitational forces and hydrodynamics are calculated on smaller time steps for the particles individually. In order to avoid very large time steps for the modified gravity solver, the global modified gravity time step is calculated according to 

\begin{align}
\Delta t_{\rm MG\, global} = \min(\Delta t_{\rm MG\, i}),
\end{align}

where $\Delta t_{\rm MG\, i}$ is the individual modified gravity time step for particle $i$ assessed using an acceleration criterion. The global PM+modified gravity time step is finally determined by

\begin{align}
\Delta t_{\rm MG+PM\, global}  = \min(\Delta t_{\rm MG\, global}, \Delta t_{\rm PM\, global} ).
\end{align}

\subsection{Power spectrum calculation}

The power spectra of the total density field and the density field of the matter components are calculated using a module implemented within the \textsc{arepo} code. The resolution elements are assigned to a Cartesian mesh using a CIC mass assignment in order to perform the required Fourier transforms \cite[see][for a detailed description]{springel2018}. 

\section*{Corresponding author}
Please address any correspondence related to this work to Christian Arnold (christian.arnold@durham.ac.uk).

\section*{Acknowledgements}
We would like to thank the Illustris-TNG collaboration for allowing us to use their baryonic model to carry out the simulations presented in this work. We are particularly grateful to Volker Springel and Rainer Weinberger for their help with the {\sc arepo} code and for useful discussions on the results, and to Alejandro Benitez-Llambay for making Py-SPHViewer \citep{llambay2015} available. Special thanks to Carlos Frenk and Jianhua He for their insightful comments on the results.

This work described in this paper is supported by the European Research Council through an ERC Starting Grant (ERC-StG-716532-PUNCA). BL is additionally supported by STFC Consolidated Grants ST/P000541/1 and ST/L00075X/1.

The cosmological simulations described in this work were run on the DiRAC Data Centric System at Durham University, United Kingdom, operated by the Institute for Computational Cosmology on behalf of the STFC DiRAC HPC Facility (www.dirac.ac.uk). This equipment was funded by BIS National E-infrastructure capital grant ST/K00042X/1, STFC capital grants ST/H008519/1 and ST/K00087X/1, STFC DiRAC Operations grant ST/K003267/1 and Durham University. DiRAC is part of the National E-Infrastructure.

The simulation code \textsc{arepo} \citep{springel2010} is currently not publicly available. The analysis scripts used to analyze the simulation output can be made available to the reader on request.

\section*{Competing interests}
The authors declare no competing interests.

\section*{Author contributions}
C.A. and B.L. planned the project. C.A. developed, implemented and optimized (together with B.L.) the modified gravity solver AREPO, ran the simulations and performed main part of the analysis.  M.L. performed the analysis for the HI power spectrum. C.A., B.L. and M.L. interpreted the results. C.A. wrote the manuscript with contributions from M.L. and B.L.

\section*{Supplementary Material}
Supplementary material and the published version of this preprint are available from this website: \url{https://www.nature.com/articles/s41550-019-0823-y}


\end{document}

%% file: JournalAbbr.tex
\def\apj{ApJ}%
\def\aap{A\&A}%
\def\mnras{MNRAS}%
\def\prd{Phys.~Rev.~D}%
\def\nat{Nature}%
\def\physrep{Phys.~Rep.}%
\def\jcap{JCAP}

%% file: second_version_arXiv.bbl
\begin{thebibliography}{10}
\expandafter\ifx\csname url\endcsname\relax
  \def\url#1{\texttt{#1}}\fi
\expandafter\ifx\csname urlprefix\endcsname\relax\def\urlprefix{URL }\fi
\providecommand{\bibinfo}[2]{#2}
\providecommand{\eprint}[2][]{\url{#2}}

\bibitem{buchdahl1970}
\bibinfo{author}{{Buchdahl}, H.~A.}
\newblock \bibinfo{title}{{Non-linear Lagrangians and cosmological theory}}.
\newblock \emph{\bibinfo{journal}{\mnras}} \textbf{\bibinfo{volume}{150}},
  \bibinfo{pages}{1} (\bibinfo{year}{1970}).

\bibitem{husa2007}
\bibinfo{author}{{Hu}, W.} \& \bibinfo{author}{{Sawicki}, I.}
\newblock \bibinfo{title}{{Models of f(R) cosmic acceleration that evade solar
  system tests}}.
\newblock \emph{\bibinfo{journal}{\prd}} \textbf{\bibinfo{volume}{76}},
  \bibinfo{pages}{064004} (\bibinfo{year}{2007}).
\newblock \eprint{0705.1158}.

\bibitem{khoury2004}
\bibinfo{author}{{Khoury}, J.} \& \bibinfo{author}{{Weltman}, A.}
\newblock \bibinfo{title}{{Chameleon cosmology}}.
\newblock \emph{\bibinfo{journal}{\prd}} \textbf{\bibinfo{volume}{69}},
  \bibinfo{pages}{044026} (\bibinfo{year}{2004}).
\newblock \eprint{arXiv:astro-ph/0309411}.

\bibitem{will2014}
\bibinfo{author}{{Will}, C.~M.}
\newblock \bibinfo{title}{{The Confrontation between General Relativity and
  Experiment}}.
\newblock \emph{\bibinfo{journal}{Living Reviews in Relativity}}
  \textbf{\bibinfo{volume}{17}}, \bibinfo{pages}{4} (\bibinfo{year}{2014}).
\newblock \eprint{1403.7377}.

\bibitem{lombriser2016}
\bibinfo{author}{{Lombriser}, L.} \& \bibinfo{author}{{Taylor}, A.}
\newblock \bibinfo{title}{{Breaking a dark degeneracy with gravitational
  waves}}.
\newblock \emph{\bibinfo{journal}{Journal of Cosmology and Astro-Particle
  Physics}} \textbf{\bibinfo{volume}{2016}}, \bibinfo{pages}{031}
  (\bibinfo{year}{2016}).

\bibitem{sakstein2017}
\bibinfo{author}{{Sakstein}, J.} \& \bibinfo{author}{{Jain}, B.}
\newblock \bibinfo{title}{{Implications of the Neutron Star Merger GW170817 for
  Cosmological Scalar-Tensor Theories}}.
\newblock \emph{\bibinfo{journal}{Physical Review Letters}}
  \textbf{\bibinfo{volume}{119}}, \bibinfo{pages}{251303}
  (\bibinfo{year}{2017}).
\newblock \eprint{1710.05893}.

\bibitem{ezquiaga2017}
\bibinfo{author}{{Ezquiaga}, J.~M.} \& \bibinfo{author}{{Zumalac{\'a}rregui},
  M.}
\newblock \bibinfo{title}{{Dark Energy After GW170817: Dead Ends and the Road
  Ahead}}.
\newblock \emph{\bibinfo{journal}{Physical Review Letters}}
  \textbf{\bibinfo{volume}{119}}, \bibinfo{pages}{251304}
  (\bibinfo{year}{2017}).

\bibitem{lombriser2017}
\bibinfo{author}{{Lombriser}, L.} \& \bibinfo{author}{{Lima}, N.~A.}
\newblock \bibinfo{title}{{Challenges to self-acceleration in modified gravity
  from gravitational waves and large-scale structure}}.
\newblock \emph{\bibinfo{journal}{Physics Letters B}}
  \textbf{\bibinfo{volume}{765}}, \bibinfo{pages}{382--385}
  (\bibinfo{year}{2017}).

\bibitem{sofa2010}
\bibinfo{author}{{Sotiriou}, T.~P.} \& \bibinfo{author}{{Faraoni}, V.}
\newblock \bibinfo{title}{{f(R) theories of gravity}}.
\newblock \emph{\bibinfo{journal}{Reviews of Modern Physics}}
  \textbf{\bibinfo{volume}{82}}, \bibinfo{pages}{451--497}
  (\bibinfo{year}{2010}).
\newblock \eprint{0805.1726}.

\bibitem{joyce2015}
\bibinfo{author}{{Joyce}, A.}, \bibinfo{author}{{Jain}, B.},
  \bibinfo{author}{{Khoury}, J.} \& \bibinfo{author}{{Trodden}, M.}
\newblock \bibinfo{title}{{Beyond the cosmological standard model}}.
\newblock \emph{\bibinfo{journal}{\physrep}} \textbf{\bibinfo{volume}{568}},
  \bibinfo{pages}{1--98} (\bibinfo{year}{2015}).
\newblock \eprint{1407.0059}.

\bibitem{schmidt2010}
\bibinfo{author}{{Schmidt}, F.}
\newblock \bibinfo{title}{{Dynamical masses in modified gravity}}.
\newblock \emph{\bibinfo{journal}{\prd}} \textbf{\bibinfo{volume}{81}},
  \bibinfo{pages}{103002} (\bibinfo{year}{2010}).
\newblock \eprint{1003.0409}.

\bibitem{zhao2011}
\bibinfo{author}{{Zhao}, G.-B.}, \bibinfo{author}{{Li}, B.} \&
  \bibinfo{author}{{Koyama}, K.}
\newblock \bibinfo{title}{{Testing Gravity Using the Environmental Dependence
  of Dark Matter Halos}}.
\newblock \emph{\bibinfo{journal}{Physical Review Letters}}
  \textbf{\bibinfo{volume}{107}}, \bibinfo{pages}{071303}
  (\bibinfo{year}{2011}).
\newblock \eprint{1105.0922}.

\bibitem{li2011}
\bibinfo{author}{{Li}, Y.} \& \bibinfo{author}{{Hu}, W.}
\newblock \bibinfo{title}{{Chameleon halo modeling in f(R) gravity}}.
\newblock \emph{\bibinfo{journal}{\prd}} \textbf{\bibinfo{volume}{84}},
  \bibinfo{pages}{084033} (\bibinfo{year}{2011}).
\newblock \eprint{1107.5120}.

\bibitem{lombriser2013}
\bibinfo{author}{{Lombriser}, L.}, \bibinfo{author}{{Li}, B.},
  \bibinfo{author}{{Koyama}, K.} \& \bibinfo{author}{{Zhao}, G.-B.}
\newblock \bibinfo{title}{{Modeling halo mass functions in chameleon f(R)
  gravity}}.
\newblock \emph{\bibinfo{journal}{\prd}} \textbf{\bibinfo{volume}{87}},
  \bibinfo{pages}{123511} (\bibinfo{year}{2013}).
\newblock \eprint{1304.6395}.

\bibitem{puchwein2013}
\bibinfo{author}{{Puchwein}, E.}, \bibinfo{author}{{Baldi}, M.} \&
  \bibinfo{author}{{Springel}, V.}
\newblock \bibinfo{title}{{Modified-Gravity-GADGET: a new code for cosmological
  hydrodynamical simulations of modified gravity models}}.
\newblock \emph{\bibinfo{journal}{\mnras}} \textbf{\bibinfo{volume}{436}},
  \bibinfo{pages}{348--360} (\bibinfo{year}{2013}).
\newblock \eprint{1305.2418}.

\bibitem{hellwing2013}
\bibinfo{author}{{Hellwing}, W.~A.}, \bibinfo{author}{{Li}, B.},
  \bibinfo{author}{{Frenk}, C.~S.} \& \bibinfo{author}{{Cole}, S.}
\newblock \bibinfo{title}{{Hierarchical clustering in chameleon f(R) gravity}}.
\newblock \emph{\bibinfo{journal}{\mnras}} \textbf{\bibinfo{volume}{435}},
  \bibinfo{pages}{2806--2821} (\bibinfo{year}{2013}).
\newblock \eprint{1305.7486}.

\bibitem{zivick2015}
\bibinfo{author}{{Zivick}, P.}, \bibinfo{author}{{Sutter}, P.~M.},
  \bibinfo{author}{{Wandelt}, B.~D.}, \bibinfo{author}{{Li}, B.} \&
  \bibinfo{author}{{Lam}, T.~Y.}
\newblock \bibinfo{title}{{Using cosmic voids to distinguish f(R) gravity in
  future galaxy surveys}}.
\newblock \emph{\bibinfo{journal}{\mnras}} \textbf{\bibinfo{volume}{451}},
  \bibinfo{pages}{4215--4222} (\bibinfo{year}{2015}).
\newblock \eprint{1411.5694}.

\bibitem{cautun2018}
\bibinfo{author}{{Cautun}, M.} \emph{et~al.}
\newblock \bibinfo{title}{{The Santiago-Harvard-Edinburgh-Durham void
  comparison I: SHEDding light on chameleon gravity tests}}.
\newblock \emph{\bibinfo{journal}{\mnras}}  (\bibinfo{year}{2018}).
\newblock \eprint{1710.01730}.

\bibitem{mitchell2018}
\bibinfo{author}{{Mitchell}, M.~A.}, \bibinfo{author}{{He}, J.-h.},
  \bibinfo{author}{{Arnold}, C.} \& \bibinfo{author}{{Li}, B.}
\newblock \bibinfo{title}{{A general framework to test gravity using galaxy
  clusters - I. Modelling the dynamical mass of haloes in f(R) gravity}}.
\newblock \emph{\bibinfo{journal}{\mnras}} \textbf{\bibinfo{volume}{477}},
  \bibinfo{pages}{1133--1152} (\bibinfo{year}{2018}).

\bibitem{li2018}
\bibinfo{author}{{Li}, B.} \& \bibinfo{author}{{Shirasaki}, M.}
\newblock \bibinfo{title}{{Galaxy-galaxy weak gravitational lensing in f(R)
  gravity}}.
\newblock \emph{\bibinfo{journal}{\mnras}} \textbf{\bibinfo{volume}{474}},
  \bibinfo{pages}{3599--3614} (\bibinfo{year}{2018}).
\newblock \eprint{1710.07291}.

\bibitem{arnold2014}
\bibinfo{author}{{Arnold}, C.}, \bibinfo{author}{{Puchwein}, E.} \&
  \bibinfo{author}{{Springel}, V.}
\newblock \bibinfo{title}{{Scaling relations and mass bias in hydrodynamical f
  (R) gravity simulations of galaxy clusters}}.
\newblock \emph{\bibinfo{journal}{\mnras}} \textbf{\bibinfo{volume}{440}},
  \bibinfo{pages}{833--842} (\bibinfo{year}{2014}).
\newblock \eprint{1311.5560}.

\bibitem{arnold2015}
\bibinfo{author}{{Arnold}, C.}, \bibinfo{author}{{Puchwein}, E.} \&
  \bibinfo{author}{{Springel}, V.}
\newblock \bibinfo{title}{{The Lyman {$\alpha$} forest in f(R) modified
  gravity}}.
\newblock \emph{\bibinfo{journal}{\mnras}} \textbf{\bibinfo{volume}{448}},
  \bibinfo{pages}{2275--2283} (\bibinfo{year}{2015}).
\newblock \eprint{1411.2600}.

\bibitem{hammami2015}
\bibinfo{author}{{Hammami}, A.}, \bibinfo{author}{{Llinares}, C.},
  \bibinfo{author}{{Mota}, D.~F.} \& \bibinfo{author}{{Winther}, H.~A.}
\newblock \bibinfo{title}{{Hydrodynamic effects in the symmetron and
  f(R)-gravity models}}.
\newblock \emph{\bibinfo{journal}{\mnras}} \textbf{\bibinfo{volume}{449}},
  \bibinfo{pages}{3635--3644} (\bibinfo{year}{2015}).
\newblock \eprint{1503.02004}.

\bibitem{he2016}
\bibinfo{author}{{He}, J.-h.} \& \bibinfo{author}{{Li}, B.}
\newblock \bibinfo{title}{{Accurate method of modeling cluster scaling
  relations in modified gravity}}.
\newblock \emph{\bibinfo{journal}{\prd}} \textbf{\bibinfo{volume}{93}},
  \bibinfo{pages}{123512} (\bibinfo{year}{2016}).
\newblock \eprint{1508.07350}.

\bibitem{winther2015}
\bibinfo{author}{{Winther}, H.~A.} \emph{et~al.}
\newblock \bibinfo{title}{{Modified gravity N-body code comparison project}}.
\newblock \emph{\bibinfo{journal}{\mnras}} \textbf{\bibinfo{volume}{454}},
  \bibinfo{pages}{4208--4234} (\bibinfo{year}{2015}).
\newblock \eprint{1506.06384}.

\bibitem{springel2010}
\bibinfo{author}{{Springel}, V.}
\newblock \bibinfo{title}{{E pur si muove: Galilean-invariant cosmological
  hydrodynamical simulations on a moving mesh}}.
\newblock \emph{\bibinfo{journal}{\mnras}} \textbf{\bibinfo{volume}{401}},
  \bibinfo{pages}{791--851} (\bibinfo{year}{2010}).
\newblock \eprint{0901.4107}.

\bibitem{pillepich2018b}
\bibinfo{author}{{Pillepich}, A.} \emph{et~al.}
\newblock \bibinfo{title}{{Simulating galaxy formation with the IllustrisTNG
  model}}.
\newblock \emph{\bibinfo{journal}{\mnras}} \textbf{\bibinfo{volume}{473}},
  \bibinfo{pages}{4077--4106} (\bibinfo{year}{2018}).

\bibitem{springel2018}
\bibinfo{author}{{Springel}, V.} \emph{et~al.}
\newblock \bibinfo{title}{{First results from the IllustrisTNG simulations:
  matter and galaxy clustering}}.
\newblock \emph{\bibinfo{journal}{\mnras}} \textbf{\bibinfo{volume}{475}},
  \bibinfo{pages}{676--698} (\bibinfo{year}{2018}).
\newblock \eprint{1707.03397}.

\bibitem{genel2018}
\bibinfo{author}{{Genel}, S.} \emph{et~al.}
\newblock \bibinfo{title}{{The size evolution of star-forming and quenched
  galaxies in the IllustrisTNG simulation}}.
\newblock \emph{\bibinfo{journal}{\mnras}} \textbf{\bibinfo{volume}{474}},
  \bibinfo{pages}{3976--3996} (\bibinfo{year}{2018}).

\bibitem{marinacci2018}
\bibinfo{author}{{Marinacci}, F.} \emph{et~al.}
\newblock \bibinfo{title}{{First results from the IllustrisTNG simulations:
  radio haloes and magnetic fields}}.
\newblock \emph{\bibinfo{journal}{\mnras}} \textbf{\bibinfo{volume}{480}},
  \bibinfo{pages}{5113--5139} (\bibinfo{year}{2018}).

\bibitem{nelson2018}
\bibinfo{author}{{Nelson}, D.} \emph{et~al.}
\newblock \bibinfo{title}{{First results from the IllustrisTNG simulations: the
  galaxy colour bimodality}}.
\newblock \emph{\bibinfo{journal}{\mnras}} \textbf{\bibinfo{volume}{475}},
  \bibinfo{pages}{624--647} (\bibinfo{year}{2018}).
\newblock \eprint{1707.03395}.

\bibitem{arnold2018}
\bibinfo{author}{{Arnold}, C.}, \bibinfo{author}{{Fosalba}, P.},
  \bibinfo{author}{{Springel}, V.}, \bibinfo{author}{{Puchwein}, E.} \&
  \bibinfo{author}{{Blot}, L.}
\newblock \bibinfo{title}{{The modified gravity lightcone simulation project I:
  Statistics of matter and halo distributions}}.
\newblock \emph{\bibinfo{journal}{\mnras}} \textbf{\bibinfo{volume}{483}},
  \bibinfo{pages}{790-805} (\bibinfo{year}{2019}).

\bibitem{ferrero2017}
\bibinfo{author}{{Ferrero}, I.} \emph{et~al.}
\newblock \bibinfo{title}{{Size matters: abundance matching, galaxy sizes, and
  the Tully-Fisher relation in EAGLE}}.
\newblock \emph{\bibinfo{journal}{\mnras}} \textbf{\bibinfo{volume}{464}},
  \bibinfo{pages}{4736--4746} (\bibinfo{year}{2017}).

\bibitem{hellwing2016}
\bibinfo{author}{{Hellwing}, W.~A.} \emph{et~al.}
\newblock \bibinfo{title}{{The effect of baryons on redshift space distortions
  and cosmic density and velocity fields in the EAGLE simulation}}.
\newblock \emph{\bibinfo{journal}{\mnras}} \textbf{\bibinfo{volume}{461}},
  \bibinfo{pages}{L11--L15} (\bibinfo{year}{2016}).

\bibitem{pillepich2018}
\bibinfo{author}{{Pillepich}, A.} \emph{et~al.}
\newblock \bibinfo{title}{{First results from the IllustrisTNG simulations: the
  stellar mass content of groups and clusters of galaxies}}.
\newblock \emph{\bibinfo{journal}{\mnras}} \textbf{\bibinfo{volume}{475}},
  \bibinfo{pages}{648--675} (\bibinfo{year}{2018}).
\newblock \eprint{1707.03406}.

\bibitem{arnold2016}
\bibinfo{author}{{Arnold}, C.}, \bibinfo{author}{{Springel}, V.} \&
  \bibinfo{author}{{Puchwein}, E.}
\newblock \bibinfo{title}{{Zoomed cosmological simulations of Milky Way-sized
  haloes in f(R) gravity}}.
\newblock \emph{\bibinfo{journal}{\mnras}} \textbf{\bibinfo{volume}{462}},
  \bibinfo{pages}{1530--1541} (\bibinfo{year}{2016}).
\newblock \eprint{1604.06095}.

\bibitem{navarro2018}
\bibinfo{author}{{Villaescusa-Navarro}, F.} \emph{et~al.}
\newblock \bibinfo{title}{{Ingredients for 21 cm Intensity Mapping}}.
\newblock \emph{\bibinfo{journal}{\apj}} \textbf{\bibinfo{volume}{866}},
  \bibinfo{pages}{135} (\bibinfo{year}{2018}).
\newblock \eprint{1804.09180}.

\bibitem{ska2015}
\bibinfo{author}{{Santos}, M.} \emph{et~al.}
\newblock \bibinfo{title}{{Cosmology from a SKA HI intensity mapping survey}}.
\newblock In \emph{\bibinfo{booktitle}{Advancing Astrophysics with the Square
  Kilometre Array (AASKA14)}}, \bibinfo{pages}{19} (\bibinfo{year}{2015}).
\newblock \eprint{1501.03989}.

\bibitem{altay2013}
\bibinfo{author}{{Altay}, G.}, \bibinfo{author}{{Theuns}, T.},
  \bibinfo{author}{{Schaye}, J.}, \bibinfo{author}{{Booth}, C.~M.} \&
  \bibinfo{author}{{Dalla Vecchia}, C.}
\newblock \bibinfo{title}{{The impact of different physical processes on the
  statistics of Lyman-limit and damped Lyman {\ensuremath{\alpha}} absorbers}}.
\newblock \emph{\bibinfo{journal}{\mnras}} \textbf{\bibinfo{volume}{436}},
  \bibinfo{pages}{2689--2707} (\bibinfo{year}{2013}).
\newblock \eprint{1307.6879}.

\bibitem{behroozi2013}
\bibinfo{author}{{Behroozi}, P.~S.}, \bibinfo{author}{{Wechsler}, R.~H.} \&
  \bibinfo{author}{{Conroy}, C.}
\newblock \bibinfo{title}{{The Average Star Formation Histories of Galaxies in
  Dark Matter Halos from z = 0-8}}.
\newblock \emph{\bibinfo{journal}{\apj}} \textbf{\bibinfo{volume}{770}},
  \bibinfo{pages}{57} (\bibinfo{year}{2013}).
\newblock \eprint{1207.6105}.

\bibitem{dsouza2015}
\bibinfo{author}{{D'Souza}, R.}, \bibinfo{author}{{Vegetti}, S.} \&
  \bibinfo{author}{{Kauffmann}, G.}
\newblock \bibinfo{title}{{The massive end of the stellar mass function}}.
\newblock \emph{\bibinfo{journal}{\mnras}} \textbf{\bibinfo{volume}{454}},
  \bibinfo{pages}{4027--4036} (\bibinfo{year}{2015}).
\newblock \eprint{1509.07418}.

\bibitem{bernardi2013}
\bibinfo{author}{{Bernardi}, M.} \emph{et~al.}
\newblock \bibinfo{title}{{The massive end of the luminosity and stellar mass
  functions: dependence on the fit to the light profile}}.
\newblock \emph{\bibinfo{journal}{\mnras}} \textbf{\bibinfo{volume}{436}},
  \bibinfo{pages}{697--704} (\bibinfo{year}{2013}).
\newblock \eprint{1304.7778}.

\bibitem{baldri2012}
\bibinfo{author}{{Baldry}, I.~K.} \emph{et~al.}
\newblock \bibinfo{title}{{Galaxy And Mass Assembly (GAMA): the galaxy stellar
  mass function at z \&lt; 0.06}}.
\newblock \emph{\bibinfo{journal}{\mnras}} \textbf{\bibinfo{volume}{421}},
  \bibinfo{pages}{621--634} (\bibinfo{year}{2012}).
\newblock \eprint{1111.5707}.

\bibitem{giodini2009}
\bibinfo{author}{{Giodini}, S.} \emph{et~al.}
\newblock \bibinfo{title}{{Stellar and Total Baryon Mass Fractions in Groups
  and Clusters Since Redshift 1}}.
\newblock \emph{\bibinfo{journal}{\apj}} \textbf{\bibinfo{volume}{703}},
  \bibinfo{pages}{982--993} (\bibinfo{year}{2009}).
\newblock \eprint{0904.0448}.

\bibitem{lovisari2015}
\bibinfo{author}{{Lovisari}, L.}, \bibinfo{author}{{Reiprich}, T.~H.} \&
  \bibinfo{author}{{Schellenberger}, G.}
\newblock \bibinfo{title}{{Scaling properties of a complete X-ray selected
  galaxy group sample}}.
\newblock \emph{\bibinfo{journal}{\aap}} \textbf{\bibinfo{volume}{573}},
  \bibinfo{pages}{A118} (\bibinfo{year}{2015}).
\newblock \eprint{1409.3845}.

\bibitem{planck2016}
\bibinfo{author}{{Planck Collaboration}} \emph{et~al.}
\newblock \bibinfo{title}{{Planck 2015 results. XIII. Cosmological
  parameters}}.
\newblock \emph{\bibinfo{journal}{\aap}} \textbf{\bibinfo{volume}{594}},
  \bibinfo{pages}{A13} (\bibinfo{year}{2016}).
\newblock \eprint{1502.01589}.

\bibitem{vogelsberger2014}
\bibinfo{author}{{Vogelsberger}, M.} \emph{et~al.}
\newblock \bibinfo{title}{{Properties of galaxies reproduced by a hydrodynamic
  simulation}}.
\newblock \emph{\bibinfo{journal}{\nat}} \textbf{\bibinfo{volume}{509}},
  \bibinfo{pages}{177--182} (\bibinfo{year}{2014}).
\newblock \eprint{1405.1418}.

\bibitem{vogelsberger2014b}
\bibinfo{author}{{Vogelsberger}, M.} \emph{et~al.}
\newblock \bibinfo{title}{{Introducing the Illustris Project: simulating the
  coevolution of dark and visible matter in the Universe}}.
\newblock \emph{\bibinfo{journal}{\mnras}} \textbf{\bibinfo{volume}{444}},
  \bibinfo{pages}{1518--1547} (\bibinfo{year}{2014}).
\newblock \eprint{1405.2921}.

\bibitem{weinberger2017}
\bibinfo{author}{{Weinberger}, R.} \emph{et~al.}
\newblock \bibinfo{title}{{Simulating galaxy formation with black hole driven
  thermal and kinetic feedback}}.
\newblock \emph{\bibinfo{journal}{\mnras}} \textbf{\bibinfo{volume}{465}},
  \bibinfo{pages}{3291--3308} (\bibinfo{year}{2017}).

\bibitem{sawicki2015}
\bibinfo{author}{{Sawicki}, I.} \& \bibinfo{author}{{Bellini}, E.}
\newblock \bibinfo{title}{{Limits of quasistatic approximation in
  modified-gravity cosmologies}}.
\newblock \emph{\bibinfo{journal}{\prd}} \textbf{\bibinfo{volume}{92}},
  \bibinfo{pages}{084061} (\bibinfo{year}{2015}).
\newblock \eprint{1503.06831}.

\bibitem{terukina2014}
\bibinfo{author}{{Terukina}, A.} \emph{et~al.}
\newblock \bibinfo{title}{{Testing chameleon gravity with the Coma cluster}}.
\newblock \emph{\bibinfo{journal}{\jcap}} \textbf{\bibinfo{volume}{4}},
  \bibinfo{pages}{013} (\bibinfo{year}{2014}).
\newblock \eprint{1312.5083}.

\bibitem{he2018}
\bibinfo{author}{{He}, J.-h.}, \bibinfo{author}{{Guzzo}, L.},
  \bibinfo{author}{{Li}, B.} \& \bibinfo{author}{{Baugh}, C.~M.}
\newblock \bibinfo{title}{{No evidence for modifications of gravity from galaxy
  motions on cosmological scales}}.
\newblock \emph{\bibinfo{journal}{Nature Astronomy}}
  \textbf{\bibinfo{volume}{2}}, \bibinfo{pages}{967--972}
  (\bibinfo{year}{2018}).
\newblock \eprint{1809.09019}.

\bibitem{springel2005c}
\bibinfo{author}{{Springel}, V.}
\newblock \bibinfo{title}{{The cosmological simulation code GADGET-2}}.
\newblock \emph{\bibinfo{journal}{\mnras}} \textbf{\bibinfo{volume}{364}},
  \bibinfo{pages}{1105--1134} (\bibinfo{year}{2005}).
\newblock \eprint{arXiv:astro-ph/0505010}.

\bibitem{li2012}
\bibinfo{author}{{Li}, B.}, \bibinfo{author}{{Zhao}, G.-B.},
  \bibinfo{author}{{Teyssier}, R.} \& \bibinfo{author}{{Koyama}, K.}
\newblock \bibinfo{title}{{ECOSMOG: an Efficient COde for Simulating MOdified
  Gravity}}.
\newblock \emph{\bibinfo{journal}{\jcap}} \textbf{\bibinfo{volume}{1}},
  \bibinfo{pages}{51} (\bibinfo{year}{2012}).
\newblock \eprint{1110.1379}.

\bibitem{bose2017}
\bibinfo{author}{{Bose}, S.} \emph{et~al.}
\newblock \bibinfo{title}{{Speeding up N-body simulations of modified gravity:
  chameleon screening models}}.
\newblock \emph{\bibinfo{journal}{Journal of Cosmology and Astro-Particle
  Physics}} \textbf{\bibinfo{volume}{2017}}, \bibinfo{pages}{050}
  (\bibinfo{year}{2017}).

\bibitem{llinares2014}
\bibinfo{author}{{Llinares}, C.}, \bibinfo{author}{{Mota}, D.~F.} \&
  \bibinfo{author}{{Winther}, H.~A.}
\newblock \bibinfo{title}{{ISIS: a new N-body cosmological code with scalar
  fields based on RAMSES. Code presentation and application to the shapes of
  clusters}}.
\newblock \emph{\bibinfo{journal}{\aap}} \textbf{\bibinfo{volume}{562}},
  \bibinfo{pages}{A78} (\bibinfo{year}{2014}).
\newblock \eprint{1307.6748}.

\bibitem{numerical}
\bibinfo{editor}{Press, W.~H.}, \bibinfo{editor}{Teukolsky, S.~A.},
  \bibinfo{editor}{Vetterling, W.~T.} \& \bibinfo{editor}{Flannery, B.~P.}
  (eds.) \emph{\bibinfo{title}{Numerical recipes}}
  (\bibinfo{publisher}{Cambridge Univ. Press}, \bibinfo{address}{Cambridge
  [u.a.]}, \bibinfo{year}{2007}), \bibinfo{edition}{3. ed.} edn.
\newblock \bibinfo{note}{Includes bibliographical references and index. - This
  printing is corrected to software version 3.0}.

\bibitem{schaye2015}
\bibinfo{author}{{Schaye}, J.} \emph{et~al.}
\newblock \bibinfo{title}{{The EAGLE project: simulating the evolution and
  assembly of galaxies and their environments}}.
\newblock \emph{\bibinfo{journal}{\mnras}} \textbf{\bibinfo{volume}{446}},
  \bibinfo{pages}{521--554} (\bibinfo{year}{2015}).
\newblock \eprint{1407.7040}.

\bibitem{llambay2015}
\bibinfo{author}{Benitez-Llambay, A.}
\newblock \bibinfo{title}{py-sphviewer: Py-sphviewer v1.0.0}
  (\bibinfo{year}{2015}).
\newblock \urlprefix\url{http://dx.doi.org/10.5281/zenodo.21703}.

\end{thebibliography}
